\documentclass{article}
\usepackage{float}
\usepackage{graphicx}
\usepackage[dvipsnames]{xcolor}
\newcommand\green[1]{{\color{Black}#1}}
\usepackage{subfiles}
\usepackage[top=2cm, bottom=2cm, left=2.5cm, right=2.5cm]{geometry}
\usepackage{titling}
\usepackage{bm}

\usepackage{dirtytalk} %nice quotation mark
\usepackage{amsmath}
\usepackage{amssymb}
\usepackage[super,numbers,sort&compress]{natbib}

\usepackage{url}
\usepackage{breakurl}
\usepackage[breaklinks]{hyperref}

\usepackage{enumitem}

\usepackage{authblk}

\usepackage{midfloat}
\usepackage{wrapfig}

\usepackage{caption}
\usepackage{setspace}

\usepackage{titlesec}

\usepackage{xfrac}

\titleformat*{\section}{\small\bfseries}
\titleformat*{\subsection}{\small\bfseries}
\usepackage{xpatch}
\usepackage{environ,needspace}

\makeatletter
\newenvironment{mytitlepage}%
  {\begin{titlepage}\def\@thanks{}}%
  {\end{titlepage}}
\xpatchcmd\titlepage{\setcounter{page}\@ne}{}{}{}
\xpatchcmd\endtitlepage{\setcounter{page}\@ne}{}{}{}
\makeatother
\begin{document}
\begin{mytitlepage}
\title{Hydrodynamic spin-orbit coupling in asynchronous optically driven micro-rotors}

\author[1,2]{Alvin Modin$^\dagger$}
\author[1,3]{Matan Yah Ben Zion$^\dagger$\thanks{Correspondence and requests for materials should be addressed to M.Y.B.Z (email:
matanbz@gmail.com)\\ $^\dagger$ Authors contributed equally.}}
\author[1]{Paul M Chaikin}

\affil[1]{\small{Center for Soft Matter Research, Department of Physics, New York University 
726 Broadway Avenue, New York, NY 10003, U.S.A.} }
\affil[2]{\small{Department of Physics and Astronomy, Johns Hopkins University, Baltimore, Maryland 21218, U.S.A.}}
\affil[3]{\small{School of Physics and Astronomy, and the Center for Physics and Chemistry of Living Systems, Tel Aviv University, Tel Aviv 6997801, Israel}}
\maketitle
\abstract{Vortical flows of rotating particles describe interactions ranging from molecular machines to atmospheric dynamics. Yet to date, direct observation of the hydrodynamic coupling between artificial micro-rotors has been restricted by the details of the chosen drive, either through synchronization (using external magnetic fields) or confinement (using optical tweezers). Here we present a new active system that illuminates the interplay of rotation and translation in free rotors. We developed a non-tweezing circularly polarized beam that simultaneously rotates hundreds of silica-coated birefringent colloids. The particles rotate asynchronously in the optical torque field while freely diffusing in the plane. We observe that neighboring particles orbit each other with an angular velocity that depends on their spins. We derive an analytical model in the Stokes limit for pairs of spheres that quantitatively explains the observed dynamics. We then find that the geometrical nature of the low Reynolds fluid results in a universal hydrodynamic spin-orbit coupling. Our findings are of significance for the understanding and development of far-from-equilibrium materials.
}

\subsection*{Introduction}

%%%%%%%%%%%%%%%%%%%%%%%%%%%%%%% Main text

% many rotors are interesting and magnetic particles are a route, but their orientations are externally imposed.
Hydrodynamically coupled rotors describe the dynamics of physically diverse systems -- the kinetics of proteins in biological membranes \cite{Davies2012}, the interactions of topological defects in superfluid Helium \cite{Stockdale2020}, and the mating rituals in dancing algae~\cite{Drescher2009}. An isotropic fluid made of rotating particles with broken parity and time reversal symmetry is expected to possess peculiar material properties, including odd viscosity~\cite{Avron1998}, and topological acoustic edge modes~\cite{Banerjee2017,Souslov2019}. Moreover, simulations show that in two-dimensional systems, hydrodynamically coupled rotors self-assemble into both random and ordered hyper-uniform arrangements~\cite{Lenz2003,Oppenheimer2019,Oppenheimer2022}.

%%%%%%%%%%%%%%%%%%% Magnetic rotors %%%%%%%%%%%%%%%
Recently, mass-produced magnetic micro-particles rotated by an external electromagnet have been extensively used as a model system for studying rotating ensembles. The rotating external field directly spins the particles, revealing new collective dynamics, including dislocation kinetics in rotating crystals~\cite{Yan2015}, propagation of chiral surface waves in a rotating liquid~\cite{Soni2019}, as well as self-healing and coarsening of colloidal fluids and crystals~\cite{MassanaCid2021}. However, magnetic rotors are not free. Just like compass needles, the orientations of the magnetic dipoles are enslaved to the globally imposed north, as they are synchronized with the orientation of the electromagnet, $\theta_M$. Though the position of the center of particle $i$, $\bm{R_i}$, may freely diffuse in the plane, the ensemble is not isotropic, as the orientational degree of freedom, $\theta_i$, is externally imposed such that $\langle \theta_i \left(t\right)\rangle\approx \theta_M\left(t\right)$. 

%$\langle \theta_i (t)\rangle \left(t\right)\approx \theta_M\left(t\right)$.

%%%%%%%%%%%%%%%%%%% Tweezed optical rotors %%%%%%%%%%%%%%%
An alternative to magnetic rotors is particles spun by a focused beam of circularly polarized light. Photonic angular momentum can be transferred to a micro-particle through its shape anisotropy \cite{Bonin2002}, birefringence \cite{Cheng2002,Bishop2004,Arita2016}, or simply by absorption \cite{Reimann2018}. Unlike magnetic rotors, photonic rotors do not directly follow the rapidly rotating electromagnetic field. Instead, the optical angular momentum flux creates a torque that maintains a steady rotation, asynchronous from the external drive. However, when using a focused beam, the position of an optical rotor, $\mathbf{R}$, is limited by the strong tweezing force at the focal point, $\mathbf{R_L}$, constraining its translational degrees of freedom $\mathbf{R}\left(t\right) \approx \mathbf{R_L}$~\cite{Ashkin1970,Ladavac2005,Shpaisman2013}. For truly \textit{free} rotors, both the orientations and the positions are free dynamic variables, and when in a liquid, $\mathbf{R}$ and $\theta$ are expected to couple hydrodynamically \cite{Faxen1922,Davis1969,Yeo2015}. %\green{Whereas observations in biological micro-rotors display hydrodynamic $\bf{R}-\theta$ coupling~\cite{Drescher2009,tan_odd_2022}, experiments with synthetic micro-rotors show incompatible dynamics \cite{MassanaCid2019,Yan2015}.}
\green{The motion of synthetic micro-rotors studied so far is incompatible with the ${\bf R}-\theta$ hydrodynamic coupling observed in pairs of biological micro-rotors~\cite{Drescher2009, MassanaCid2019}. Moreover, previous studies with ensembles of synthetic micro-rotors\cite{Yan2015,Soni2019} spin by an externally imposed field and can not show spontaneous symmetry breaking as seen in ensembles of biological rotors~\cite{tan_odd_2022}.}
To date, configuration space is reduced in either magnetic or photonic rotors, and the nature of the chosen drive obscures mutual hydrodynamic coupling. 

%%%%%%%%%%%%%%%%%%% In this Communication - free rotors %%%%%%%%%%%%%%%

In this communication, we show that optically driven rotors in a non-tweezing beam freely diffuse while spinning asynchronously. By developing a novel experimental test bed that drives hundreds of free micro-rotors (Fig.~\ref{fig:Async} a,b and Supplementary Video 1), we measure their stochastic translational and rotational dynamics independently. We find that in this system, remote particles are rotating asynchronously, and at close proximity, rotation and translation couple and rotor pairs mutually advect into an orbital motion (Supplementary Video 2). We observe that the translation and rotation of these free optical rotors reciprocate --- their spin coupling obeys a geometrical relation following the Stokes flow of spheres near a wall. To create a system of asynchronous rotors, we designed an optical setup capable of producing a uniform torque field with minimal tweezing. We also developed a synthetic route for stable silica-coated birefringent particles (Fig.~\ref{fig:Synthesis}) that rotate in a circularly polarized collimated beam. We characterize the translational and rotational motion of individual particles and pairs of particles and derive an analytical hydrodynamic model that quantitatively captures their dynamics.

\subsection*{Results and Discussion}

\subsubsection*{Synthesis of stable birefringent micro-particles}

We couple photonic angular momentum to the particles by synthesizing a new type of birefringent colloid made of silica-coated vaterite. Vaterite has a hexagonal symmetry with a positive uni-axial optical response and birefringence of $\Delta n = n_e - n_0 = 0.1$, where $n_{o} = 1.55$ and $n_e = 1.65$ are the refractive indices along the ordinary and extraordinary axes. When illuminated with circularly polarized light, vaterite particles begin to rotate while experiencing negligible thermal absorption \cite{Arita2016}, making high particle concentrations experimentally accessible without overheating (Fig.~\ref{fig:Async}b). Using previous synthetic routes~\cite{Parkin2009,Vogel2009}, we found colloidal vaterite's rotational dynamics to be inconsistent between tweezed and tweezing-free optical fields. In the absence of tweezing, particle rotation was intermittent and gradually diminished, suggesting that minor surface chemistry variations dominated the dynamics. We developed an alternative synthetic strategy to allow consistent rotation over prolonged durations. 

Particles were synthesized via controlled precipitation of highly concentrated solutions of calcium chloride $\rm{CaCl_{2}}$ and sodium carbonate $\rm{Na_{2} C.O._{3}}$. 
\begin{equation}
\rm{CaCl_{2}{}_{(aq)}} + \rm{Na_{2}CO_{3}{}_{(aq)}}\rightarrow \rm{CaCO_{3}{}_{(s)}}+\rm{2NaCl{}_{(aq)}}
\label{eqPrecipitation}
\end{equation}
The vaterite phase, a product of Eq.~\ref{eqPrecipitation}, is a metastable polymorph of calcium carbonate. Deviations towards even weakly acidic conditions cause rapid dissolution and transformation of vaterite spheres to calcite cubes \cite{Kralj1990}. Therefore, the vaterite-to-calcite phase transition is a significant barrier encountered when synthesizing and re-suspending vaterite microspheres. 

To preserve micro-particles in the vaterite phase, we control the pH of their solution and repeatedly coat particles with silica (Fig. ~\ref{fig:Synthesis}). Synthesis solutions were buffered to $pH = 9.5$ by n-cyclohexyl-2-aminoethanesulfonic (CHES) acid (see Methods for details). A typical yield for these conditions results in a cloudy suspension of billions of particles with a mean size of $3.5 \pm 0.8 ~\mu$m (see Supplementary Information, Fig. S2 a,b). We control the particles' size by varying the stirring speed, initial reactant concentration, and reaction time \cite{Volodkin2012}. To further promote phase stability in solution, limit flocculation, and preserve vaterite's optomechanical behavior, particles are coated with silica by the addition of (3-aminopropyl)trimethoxysilane (APTMS) followed by tetraethyl orthosilicate (TEOS, see Methods and Fig.\hspace{-1mm}~\ref{fig:Synthesis} (2,3)~\cite{Vogel2009}). Repeated silica precipitation alters the particles to be inert with long-term stability at room temperature. Typical experiments were performed in heavy water ($\rm{D}_2\rm{O}$), selected for its lower absorption of infrared radiation.

\subsubsection*{Translational dynamics of individual rotors}

When the laser is turned off, vaterite micro-spheres sediment onto the glass surface ($\rho_{\rm{vaterite}} = 2.54\;\rm{g/cc}$) and diffuse in a quasi-two-dimensional plane. The gravitational height is $h_g= k_B T/F_{\rm{g}} \approx 3 - 20\;\rm{nm} $, (where $T$ is the absolute temperature, and $F_{\rm{g}} = \pi g d^3 \Delta \rho$/6 is the buoyant force given by the buoyant density $\Delta \rho = \rho_{\rm{vaterite}}-\rho_{\rm{D}_2\rm{O}}$), for particles within the examined size range ($d \approx 3-6\;\mu\rm{m}$). \green{Measuring the particles' mean squared displacements (MSDs), $\langle \Delta r^2\rangle\green{=4D_t \tau}$, we observe a reduction in their translational diffusion constants $D_{t}$ compared to the bulk value, $D^{\rm{bulk}}_t=k_B T/3\pi \eta d$ (Fig.~\ref{fig:TransDyn}b)} \cite{allan_daniel_2016_60550,crocker_methods_1996}. For each particle, the mean proximity to the surface, $h$, is given by the gravitational height, and in the lubrication limit ($\langle h\rangle=h_g \ll d$) reduces the translational mobility \cite{Goldman1967}. The expected  translational diffusion coefficient is
\begin{equation}
    D_{t} \approx \frac{5k_BT}{8\pi\eta d\log(d/2h)}. 
    \label{eqDtWall}
\end{equation}

To create a tweezing-free optical torque field, we used a collimated, circularly polarized infrared (I.R.) laser ($\lambda = 1064\;\rm{nm}$). When incident onto the sample, the wide-field $D \approx 440\;\mu \rm{m}$ spot can deliver a maximum power flux, $J$, of up to $40$ MW/$m^2$. 
The flux varies spatially by less than $\pm 10\%$ within the field of view  (165$\;\mu$m x 125$\;\mu$m), assuring that our collimated beam is free of sharp intensity gradients otherwise found in focused beams (see Supplementary Information, Fig. S1b)~\cite{BenZion2022}. \green{When gradients in the beam's intensity are present, particles are constrained to the narrow waist of the focused light. A narrow-waisted beam generates tweezing forces that typically restrict the translational motion of a particle to roughly its diameter, obscuring the coupling between translation and rotation. To date, focused light beams used to rotate particles were too tight to host an ensemble of particles, making hydrodynamic particle-particle interactions inaccessible \cite{Parkin2009,molloy_lights_2002}.} 

Our setup generates a photonic torque that drives particles to rotate while freely diffusing in-plane (Figs.~\ref{fig:Async} a,b). In contrast to previous systems relying on optical tweezing or trapping, \cite{di_leonardo_hydrodynamic_2012,Bishop2004} the translational MSD. of rotating particles remains linear in time, $\langle \Delta r^2\rangle \propto \tau^1$ (Fig.\ref{fig:RotDyn}a), indicating that the broad beam profile has minimal transverse tweezing. The translational diffusion constant, $D_t$, is enhanced at higher fluxes. (Fig.\ref{fig:RotDyn}b). Radiation pressure from back-scattered photons generates a force \green{$F_{rad}$} opposite to gravity and smaller in magnitude. The force from the radiation pressure opposes the gravitational pull, $F_{g}$, effectively increasing the gravitational height to $h_{g} = k_{B}T/\left(F_g - F_{rad}\right)$. The increase in $h_{g}$ reduces the wall's effect on the drag coefficient, increasing the translational diffusion constant until it approaches its bulk value.

\green{At higher fluxes where $F_{rad}$ exceeds $F_{g}$, vaterite particles begin to steadily rise from the capillary's bottom surface at a constant speed. Vertically shifting the imaging focal plane from the bottom of the capillary to its top surface, we monitor the time it takes for particles to travel $100 \mu$m, corresponding to when a focused image of a particle re-appears (see Supplementary Information for additional experimental details).} We measure the force from radiation pressure $\langle F_{rad}\rangle = R J \pi d{^2}/4c$ by computing the particles' vertical rising speeds and extracting the intrinsic reflection coefficient from radiation pressure, $R = 0.22\pm 0.01$ (see Supplementary Information and Fig.S3). For a power flux of $40$ MW/$m^2$ incident on a $d \approx3\;\mu m$ vaterite particle, the gravitational height will increase from $\approx$ 15 nm to $\approx$ 180 nm resulting in $\approx$ 90\% increase in the translational diffusion, $D_{t}$, consistent with measured diffusion coefficient as extracted from the MSD. (Fig.~\ref{fig:RotDyn}a,b).

\subsubsection*{Rotational dynamics of individual rotors}

We measure the stochastic rotational diffusion with no drive by monitoring the transmitted light intensity of each particle. While confined to a two-dimensional plane, vaterite particles undergo rotational diffusion. When imaged under crossed-polarizers (PA), the particles' birefringence modulates the intensity of the scattered light. The fluctuations of the transmitted light intensity lead to temporal decorrelation, $g_{PA}(\tau)$, that holds information about the particle's orientational diffusion. The decorrelation rate depends on the rotational diffusion matrix $\mathbf{D}_r$, which near a wall is dominated by the spinning diffusion (rotation perpendicular to the wall), $g_{P.A.}=\exp(-6\mathbf{D}_{r}\tau)\approx \exp(-6D_{r,\perp}\tau)$ (Fig.~\ref{fig:TransDyn}c, inset and Supplementary Information for details) \cite{Berne2000, Lisicki2014}. To leading order, the diffusive spinning approaches its bulk value
\begin{equation} 
    D_{r,\perp} \approx  \frac{k{_B}T}{\pi\eta d^3},
    \label{eqDrWall}
\end{equation}
and is consistent with the experimentally measured diffusion constants in the particle size range studied (Fig.~\ref{fig:TransDyn}c). The lubrication flows responsible for the reduction in translational diffusion (\ref{eqDtWall}) have little effect on the particle's spinning relative to their bulk dynamics (\ref{eqDrWall}). This relation is significant for the spin-orbit coupling of a pair of rotors.

When illuminated with circularly polarized light, individual particles rotate at a steady angular speed of up to $d\theta/dt\equiv\Omega \approx 1.2\;\text{rad}/s$. Diffusive spinning is dominated by the external drive, with the average rotational Péclet, $\textrm{Pe}^r= \Omega/D_r\approx 25 $. For a given particle, the spinning angular frequency is given by the balance of optical torque and viscous drag 
    \begin{equation}
        \Omega=\frac{\rm{T} J \lambda}{8\pi c\eta d}\left(1-\cos\left(\frac{2\pi\Delta n d}{\lambda}\right)\right) 
        \label{eqRotsFreq} 
    \end{equation}  
where $\eta = 1.25$ m Pa$\cdot$s is the surrounding fluid's viscosity~\cite{Hardy1949},  $c$ is the speed of light in vacuum, and $\rm{T}$ is the transmission coefficient \cite{Friese1996,Vaippully2020} (\green{see Supplementary Information for detailed derivation}). In the chosen wavelength, vaterite has negligible absorption ~\cite{Arita2016}, and in the short wavelength approximation, the transmission of the refracted rays is given by $\rm{T} \approx  1 - \rm{R}$. 
%%%%Birefringence measurement%%
We measure the birefringence of the particles, $\Delta n$, by tuning the ellipticity of the incident beam. In the presence of polarized light with ellipticity $\phi$, the optical torque is composed of both an aligning and spinning torque. For circularly polarized light, the alignment torque vanishes. Conversely, the spinning torque vanishes for linearly polarized light. The effective birefringence of the polycrystalline vaterite colloids is then measured by considering the minimum ellipticity required to generate a net torque that overcomes the viscous torque $\tau_{v}$ for a given particle size (Fig.~\ref{fig:RotDyn}c). We measure the effective birefringence as $\Delta n = 0.075\pm 0.015$, consistent with $\Delta n$ values for polycrystalline vaterite micro-particles reported in the literature between $0.06-0.09$~\cite{Parkin2009,Vogel2009,Xie2021}. Using the measured birefringence, $\Delta n$, and transmission, $\rm{T}$, we quantitatively predict the rotation rate of individual particles as a function of flux $J$ and size $d$ (Fig.~\ref{fig:RotDyn}d). Note that the dependence of $\Omega$ on $d$ is non-monotonic due to the effectiveness of a particle as a wave plate, peaking at the thickness of an ideal half-wave plate $d_{\rm{max}}\approx \frac{1}{2}\frac{\lambda}{\Delta n}\approx 5\;\mu$m (see Eq.~\ref{eqRotsFreq}, and Fig.~\ref{fig:RotDyn}d).

\subsubsection*{Asynchronous rotation}

% We find particles rotate asynchronously from the independent phases of of their transmitted light. When diffusing, vaterite particles de-polarize the linearly polarized light, allowing light through the analyzer to reach the camera. 

We find that particles spin asynchronously by using microscopic imaging that monitors the intensity of light transmitted through individual vaterite microspheres. Particles are imaged between crossed-polarizers using a custom-built microscope with bright field ($\lambda = 505$ nm) illumination (see Fig.~\ref{fig:Async}a and Supplementary Information). When spinning, vaterite de-polarizes the transmitted light from the LED source periodically. This occurs whenever the optical axis of the particle coincides with neither the axis of the polarizer nor the analyzer, corresponding to four depolarizations, or "blinks'', per period. (Fig.~\ref{fig:Async}c).  \green{The relative orientation of
an ensemble of vaterite particles is free to vary.} Tracking the transmitted light intensities of individual particles as a function of time, $I_{i}\left(t\right)$, allows for a direct measure of each particle's rotation frequency and phase. Simultaneously monitoring two spinning particles shows that the periodically oscillating intensities of the light they transmit are close in frequency but differ in phase (Fig.~\ref{fig:Async}c). Computing the Fourier transform of the light intensities of individual rotors, $\mathcal{F}_{i}\left[I_{i}\left(t\right)\right]\left(\omega\right)\equiv \int dt e^{-\mathrm{i} \omega t} I_{i}\left(t\right)$, allows us to globally compare the phases of multiple particles (see Supplementary Information for details). For individual particles, the magnitude of the Fourier transform, $\sqrt{\mathcal{F}_{i}\mathcal{F}_{i}^{*}}$ peaks at $\sim 0.5\rm{ Hz}$, corresponding to four times the typical particle spinning frequency ($\sim 0.125\rm{ Hz}$). However, the sum of the individual Fourier transforms, $\left|\sum_i \mathcal{F}_i \right|^2$, decays with the number of particles. \green{The different phases of the light intensities do not necessarily add up constructively}, indicating that particles are globally asynchronous with respect to one another (Fig.~\ref{fig:Async}d). Yet when two rotating particles approach each other, they mutually advect through their flow fields. Moreover, we observe a change in their blinking rate, indicating a change in their angular speed.

\subsubsection*{Single particle flow field}
To understand the coupling of rotor pairs, we first consider the flow field generated by a single rotor. In a uniform optical torque field, an isolated spinning vaterite micro-sphere stirs the surrounding fluid, generating an algebraically decaying flow. A single rotor can be modeled as an isolated sphere with radius $a$ centered at $\bm{r}$, subjected to a constant torque $\bm{\tau} = 8\pi\eta a^3\bm{\Omega^0}$, where $\bm{\Omega^0}$ is the angular velocity of the isolated rotor (Fig.~\ref{fig:SingleRotor}). A multipole expansion well approximates the resulting flow generated by the sphere's rotation. \green{We introduce a singularity at position $\left(x_{0},y_{0},z_{0}\right)$, acting as a point-torque disturbance (rotlet). The corresponding Green's function, $\bm{G_{ij}}$, satisfying the Stoke's equations is $\bm{G_{ij}}=\frac{\epsilon_{ijk}r_{k}}{r^{3}}$~\cite{Blake1974}, where $\epsilon_{ijk}$ is the Levi-Cevita symbol whose indices represent components of the rotlet's position in the Cartesian coordinate system, and $r$ is a 3D vector pointing from the rotlet to a point $\left(x,y,z\right)$ in space}. In an unbounded 3D Stokes fluid, the magnitude of the far-field flow of a force monopole (a Stokeslet) is $\left|\bm{u}^{\rm{bulk}}_{\rm{Stokelet}}\right|\propto 1/r$, and a force dipole $\left|\bm{u}^{\rm{bulk}}_{\rm{Dipole}}\right|\propto 1/r^2$. To leading order, the rotors are force-free but experience a torque. The resulting flow is given by an anti-symmetric derivative of the Stokeslet (a rotlet) that decays as $\left|{\bm u}^{\rm{bulk}}_{\rm{rotlet}}\right|\propto 1/r^2$. 

In all experiments, the micro-rotors are found near a solid wall, imposing a no-slip boundary condition ($\bm{u}\left(z=0\right)=0$) Fig.~\ref{fig:SingleRotor}). Their flow field obtained by the method of images decays like the following derivative $ \sim 1/r^3$ ~\cite{Blake1974}. Explicitly, we may approximate the far-field fluid flow in the presence of a no-slip boundary as the superposition of two rotlets with equal and opposite torques $\bm{\tau}=\pm\bm{\tau^0}$ located at $z=\pm\delta$, respectively (Fig.~\ref{fig:SingleRotor}). The resulting fluid flow is then $\mathbf{u}\left(r\right)=\Omega^{0}a^{3}\left\{\left(\frac{1}{R_{+}^{3}}-\frac{1}{R_{-}^{3}}\right)\left(-y\mathbf{\hat{x}}+x\mathbf{\hat{y}}\right)\right\}$ \green{(see Supplementary Information for detailed calculation)}. \green{Here $\bm{\hat{x}},\bm{\hat{y}}$ are Cartesian unit vectors, and $|R_{\pm}|\equiv \left(x^2+y^2+\left(z\mp\delta\right)^{2}\right)^\frac{1}{2}$, representing the distance to a point $\left(x,y,z\right)$ in space from the source and image charges, respectively. In the far-field limit, $|R_{\pm}|^{-3}\approx\frac{1}{r^{3}}\left(1\pm\frac{3\delta^2}{r^{2}}\right)$.} When close to the wall ($\delta\approx a$), the $1/r^3$ contribution vanishes, and the next term in the multipole expansion is now proportional to $1/r^{4}$. \green{This scaling arises from noting that $-y\bm{\hat{x}}+x\bm{\hat{y}} = r\bm{\hat{\theta}}$.}  To leading order, a sphere spinning at an angular frequency $\Omega$ near a wall generates a flow of,

\begin{equation}
    {\bm u(r) } = \frac{6\Omega^0 a^3 \delta^2}{r^4} \bm{\hat{\theta}},
    \label{eqTorqueletWall}
\end{equation}
\green{where $r$ is now the distance in the two-dimensional plane}. The reciprocal motion of a pair of rotors can be described as though each generates a flow field described by Eq.~\ref{eqTorqueletWall} while being advected by the same flow profile generated by the companion rotor.

\subsubsection*{Hydrodynamic spin-orbit coupling in a pair of micro-rotors}

% Just as active Brownian particles are dubbed to have an ``active speed'' \cite{bechinger_active_2016}, active rotors can be described by an internal ``active spin''. For the low Reynolds number flows experienced by vaterite particles $\left(Re = \rho d v/\eta \approx 10^{-7}\right)$, the active spin $\mathbf{\Omega}^{0}$ is fueled by a constant external photonic torque given by $\bm{\tau^0}=8\pi\eta a^{3}\mathbf{\Omega}^{0}$,
% where $\eta$ is the fluid viscosity and $a$ is the particle radius.

When two micro-particles are sufficiently close to each other, their long-range flow fields drive them into a short-lived orbit until they diffuse apart. In a typical interaction, a pair of particles with a mean size ratio $0.75-1$, exhibit relative orbital motions for $20-30$ s before separating (Fig.~\ref{fig:SpinOrbit} and Supplementary Video 2). As two rotors orbit each other, each particle's angular speed, $\Omega$, is reduced from its nominal value (Fig.~\ref{fig:SpinOrbit}c, inset). When in close proximity, the blinking rate of the particles can decrease by up to 35\%, corresponding to a slowdown in their spinning rate. Yet particles continue to rotate asynchronously even at near contact (Supplementary Video 2). Being of finite size, a particle's rotation rate changes when subjected to flow gradients. The orbital interaction of the synthetic photonic rotors resembles the dynamics seen in \green{microscopic organisms}~\cite{Drescher2009, Huang2021,tan_odd_2022}, and stands in stark contrast to interactions seen in magnetically rotated particles \cite{MassanaCid2021}. 

The dynamics of this spin-orbit interaction can be described with a minimal far-field hydrodynamic model \green{(see Supplementary Information for detailed calculation)}. For pairs of particles $i$ and $j$ (Fig.\ref{fig:SpinOrbit}a), the velocity of a spherical particle $i$, ${\bm v_i}$ found within the flow field of particle $j$, $\bm{u}_j$ is given by Faxen's first  law~\cite{Russel1989,kim_microhydrodynamics:_2005}
    \begin{equation}
        \bm{F}_i = 6\pi \eta a\left\{\left[\bm{u}_{j}\left(\bm{r}\right)+\frac{1}{6}a^2\nabla ^2 \bm{u}_j\left(\bm{r}\right)\right]_{\bm{r}=\bm{r}_i}-\bm{v}_i\right\}, 
    \label{eqFaxen1}
    \end{equation}
where force $\bm{F}_i$ is the external force. In our case, particles are force-free $\bm{F}_i=0$ and $\nabla^2 \bm{u}$ is vanishing, so to leading order particle $i$ is simply advected by particle $j$: $\bm{v}_i \approx \bm{u}_j\left(\bm{r}=\bm{r}_i\right)$. 
 The measured orbital frequency $\bm{\omega}_{i} \equiv \bm{v}_{i}/r$ as a function of particle separation $\bm{r}$ for a pair of interacting particles is then quantitatively captured by combining Eqs. \ref{eqTorqueletWall} and \ref{eqFaxen1} (Fig. \ref{fig:SpinOrbit}c).
It is further known that the flow-induced angular velocity is proportional to the vorticity. For particle $i$, the expected change in the rotation rate, $\Delta \Omega_i$, caused by the flow generated by particle $j$ is given by Faxen's second law~\cite{Russel1989}, connecting the apparent spinning of particle $i$, $\bm{\Omega}_i$ with the flow generated by particle $j$, $\bm {u}_j$
\begin{equation}
    \bm{\Delta \Omega}_i \equiv \bm{\Omega}_i - \bm{\Omega}^0_i = \frac{1}{2}\nabla \times \bm{u}_j\left(\bm{\bm{r}=\bm{r}_i}\right).
    \label{eqFaxen2}
\end{equation} 
%%%%%%%%%%%%%%%%% Spin orbit follows a geometrical relation  %%%%%%%%%%%%%%%%
For identical rotors, $\bm{\omega}=2\bm{\omega}_{i}$ is the orbital frequency of the rotating pair about their common center. Recalling that rotating particles generate an algebraically decaying tangential flow field ($\bm{u}\propto 1/r^\alpha\bm{\hat{\theta}}$), Eq.~\ref{eqFaxen2} becomes
\begin{equation}
\bm{\Delta\Omega}=\frac{1}{4}\left(1-\alpha\right)\bm{\omega},
\label{eqSpinOrbit}
\end{equation}
connecting the spin angular frequency change to the rotating pair's orbital frequency. \green{This relation is general -- independent of $a$, $r$, and $\delta$. Every translation is accompanied by a proportional amount of rotation. Eq.~\ref{eqSpinOrbit} holds regardless of whether the flow is three-dimensional (in bulk), quasi-two-dimensional (near a wall), or strictly two-dimensional (in a liquid film).} As expected from Stokes flow, spin-orbit coupling is geometrical in nature and does not depend on physical parameters, such as the applied torque, fluid viscosity, and material composition. Surprisingly, the far-field approximation quantitatively captures the orbiting dynamics even when particles are nearly in contact. In our case, $\alpha =4$ (particles are near a wall), and as seen in Fig.~\ref{fig:SpinOrbit}d, this relation captures the spin-orbit dynamics of optical rotors of different sizes, spinning rates, and over a range of separations. This connection was not observed in previous systems of synthetic rotors and was impossible to extract from pairs of magnetic rotors which were shown to rotate as a solid body for any rotational frequency \cite{MassanaCid2021}. 

%%%%%%%%%%%%%%%%%%% Conclusions and outlook %%%%%%%%%%%%%%%%

\subsection*{Conclusions}
In this work, we introduced a new system of active spinning particles --- asynchronous photonic rotors enabled by a tweezing-free optical field. We designed a force-free torque field using a collimated beam of circularly polarized light and developed a synthetic route for birefringent silica-coated vaterite colloids to show for the first time the spinning of hundreds of micro-particles using photonic angular momentum. \green{We systematically quantified the micro-rotors' optical and hydrodynamic properties and found that particles rotate asynchronously, unlike any previous synthetic micro-rotor system. The particles' asynchronous rotation indicates that their orientational degrees of freedom are dynamic variables; this is in contrast to magnetic rotors, whose orientational degrees of freedom are "frozen" by an applied magnetic field.}
We analyzed particle spinning rate and found that pairs of rotating particles mutually advect one another, with their translation and rotation coupled hydrodynamically. Our analysis shows that, as the coupling is geometric, it may be applicable more generally in active systems, from living organisms \cite{Drescher2009,tan_odd_2022} to robotic systems \cite{scholz_rotating_2018, BenZion2023}, where translation and rotation are coupled. Our system allows for further investigation into isotropic rotating ensembles with broken time-reversal symmetry and parity, shedding light on new material properties theoretically predicted in active matter such as odd viscosity and quantum hall fluids~\cite{Avron1998,Souslov2019,goto_purely_2015}. Using non-spherical particles, free optical rotors can also be used to study the effect of morphology and steric interactions in tandem with hydrodynamic coupling~\cite{Hueckel2021a, BenZion2020}. \green{Moreover, combining our system of optical rotors with rotors driven by an external magnetic field could enable the experimental study of ensembles of counter-rotating particles, where optical rotors rotate independently from the magnetic rotors. Experimental investigation of an ensemble of counter-rotors would elucidate recent predictions on self-assembly, phase separation, and edge modes, expanding our understanding of far-from-equilibrium states of matter \cite{Yeo2015,Nguyen2014, lushi_periodic_2015,kokot_active_2017}.}
\\
 
%#################### Methods ######################
\vspace{-0.2cm}
\section*{Methods}\label{supMethods}
\label{supSamplePreparation}

%%%%%%%%%%%%%%%%%%
\subsection*{Vaterite Synthesis and Sample Preparation}
\subsubsection*{Synthesis Overview}
Vaterite micro-spheres were synthesized by controlled precipitation from a super-saturated solution of 0.33M calcium chloride ($\rm{CaCl}_{2}$, Sigma-Aldrich) and 0.33M sodium carbonate ($\rm{Na}_{2}\rm{C.O.}_{3}$, Sigma-Aldrich). We buffer $\rm{CaCl}_{2}$ and $\rm{Na}_{2}\rm{C.O.}_{3}$ to a pH of 9.5 (CHES, Sigma-Aldrich), before mixing at 1000 RPM in a glass vial with a 1 cm magnetic stir bar (Fig.~\ref{fig:Synthesis} (a, (1)). The total stirring time was approximately 40 seconds. For these conditions, the typical poly-dispersity is $3.6\pm 0.8 \mu$m, although, by varying the synthesis conditions, a particle diameter range of $d = 2-12 \mu$m is readily accessible (see Supplementary Information, Fig.S2(b)). Particles are then coated via a sequential coating process using (3-Aminopropyl)trimethoxysilane (APTMS, Sigma-Aldrich) and tetraethyl orthosilicate (TEOS, Sigma-Aldrich) (Fig.~\ref{fig:Synthesis} (a, (2) and (3)). In a typical APTMS coating, 1.5 mL of the synthesis bath is washed in DI H$_2$O 3 times, to which 70 $\mu$L of APTMS (Sigma-Aldrich), 25 $\mu$L of Ammonia (25\% V/V in $H_{2}$O, Merck) and 940 $\mu$L of ethanol (200 proof) are added. The sample is then placed in a shaker for 2.5 hours. For TEOS coatings, the procedure is identical (APTMS is replaced with TEOS), except that the sample is allowed to shake for 5.5 hours. Electrostatic interactions were minimized by the presence of 14 mM NaCl in the solution, reducing the Debye screening length to 2.5 nm. Microscope samples are made by dispersing the particles in heavy water ($\rm{D}_2\rm{O}$, Sigma-Aldrich) and loading into a $100\;\mu \rm{m}$ tall glass channel (Vitrotubes W5010050) passivated through vapor deposition of hexamethyldisilazane (Sigma-Aldrich). Loaded capillaries were placed on a clean microscope glass slide and sealed on their ends with UV-curable resin (Loon Outdoors U.V. Clear Fly Finish).

\section*{Experimental Setup}\label{supExperimentalSetup}
Imaging was done on a custom-built, bright-field microscope coupled to a laser source. A commercial light emitting diode ($\lambda = 505$ nm Thorlabs) with a diffuser (ground glass N-BK7 600 grit, Thorlabs), condenser, and an iris were used to achieve Köhler illumination. The scattered light was picked up by the microscope objective (HCX PL APO 40x NA = 0.85, Leica) and a tube lens (B\&H), detected by a digital camera (DCC1545M, Imaging Source), and acquired using commercial video recording software (I.C. Capture, Imaging Source). A laser beam was introduced on a separate optical path (Supplementary Information, Fig.S1(a)). A $\lambda = 1064$ nm laser beam (YLR-10-1064-LP, I.P.G. Photonics) was passed through a zero-order half-wave plate (WPH05M-1064 Thorlabs) and contracted using a customized Galilean telescope to achieve a wide beam (Supplementary Information, Fig.S1(b)). The laser beam was introduced into the sample using a polarizing beam splitter (PBS CM1-PBS253 Thorlabs). Its intensity at the sample was controlled by a combination of the electronic laser head controller and adjusting the half-plate. The intensity was measured using an optical power meter (PM100D power meter, with S175C sensor, Thorlabs). In order to eliminate laser intensity before the camera, stained glasses (FGS900S, Thorlabs) were stacked after the objective.

%%%% VIDEOS
%\subsection*{Supplementary Videos}
%\linespread{1}
%\begin{description}[itemsep=1.5mm]
%\item \href{https://youtu.be/ZEfo46HDgSU}{S1 - Swimmer's activity inside the dense phase.} %(./analysis/speedConcentration/animSwimbysAll.ipynb)
%\item \href{https://youtu.be/Lrz_xtDfMm4}{S2 - Entrainment mediated corralling.} %(./analysis/activityInducedPhaseSeparation/animCorraling.ipynb)
%\item \href{https://youtu.be/RI0CI5ESrjU}{S3 - Quick dispersing at low density.}
% (./videos/S3isotropicOnOff/animIsotropicOnOff.ipynb)  
%\item \href{https://youtu.be/7XR24PIX0hI}{S4 - Schooling and anti-schooler at intermediate density.}
%(./videos/S4schoolAntiSchool/plotFlocking.ipynb)
%\item \href{https://youtu.be/j6dBq0F12nY}{S5 - Internal flow inside the dense phase.}
%(./videos/S5MIPSinternalFlow/animMIPSinternalFlow.ipynb)
%\item \href{https://youtu.be/ZYLvVHBA0KY}{S6 - Corralling of passive particles.}
%(./videos/S6aipsLargeScale/animAIPS.ipynb)
%\end{description}

%################### Online Content #################
\subsubsection*{Data availability}
 Data is available at an online repository~\cite{modinFigshare2023} in the following URL: \href{http://doi.org/10.6084/m9.figshare.22294690}{http://doi.org/10.6084/m9.figshare.22294690}.

%################### Code Availability #################
\subsubsection*{Code availability}
The custom codes used in this study are available from the corresponding author upon request.

\vspace{-0.2cm}

%%%%% ACKNOWLEDGMENTS

\subsection*{Acknowledgements}
We greatly acknowledge insights and assistance from Joon Oh, Naomi Oppenheimer, Yoav Lahini, Nathaniel Spilka, Yasuo Oda, Bastian Pradenas, David Rivas, and Yihao Chen. This research was supported by the Department of Energy DE-SC0007991 for initiation and design by P.M.C. and by DOE SC0020976 for sample preparation and imaging by M.Y.B.Z. and A.M.

%%%%% Author Contributions

\subsection*{Author Contributions}
A.M., M.Y.B.Z., and P.M.C. conceived the project. A.M. and M.Y.B.Z. designed and conducted the experiments, data analysis, and developed the theoretical model. All authors contributed to the writing of the manuscript.

\subsection*{Competing interests}
The authors declare no competing interests.
\end{mytitlepage}

\pagebreak

\subsection*{Figures}
%########### FIGURE1 ############

\begin{figure}[h!]
\centering
%\vspace{0.5cm} 
\includegraphics[width=0.9\textwidth]{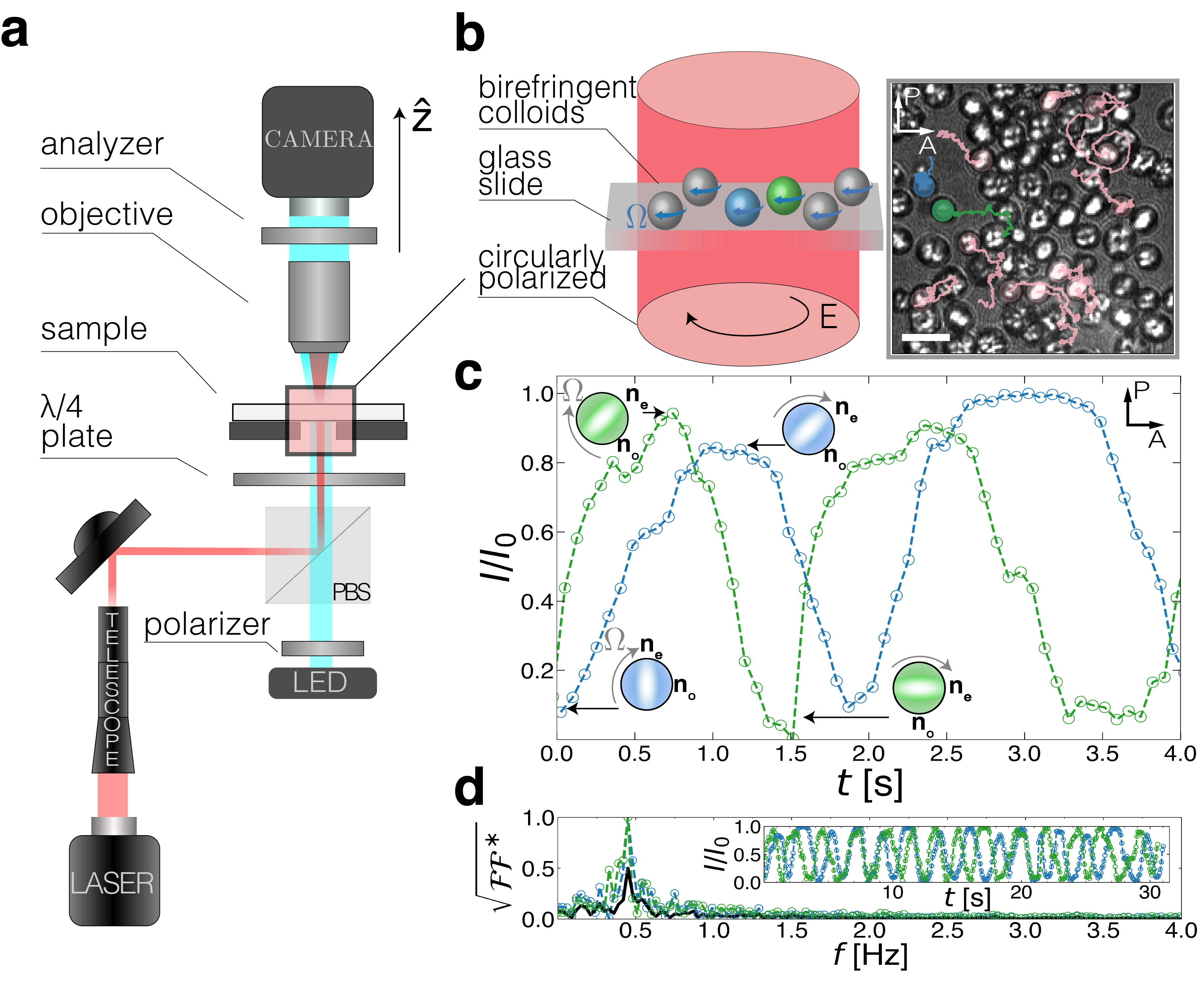} 
\caption{\textbf{Experimental setup to drive micro-rotors asynchronously.} {\bf a} Optical setup for introducing a broad ($D \approx 440\;\mu\rm{m}$) circularly polarized beam into a microscope sample. {\bf b} Schematic and polarized microscopy image of birefringent vaterite particles rotating while moving freely in the illuminated region. \green{The transmitted light intensities of two particles (blue and green) are tracked over the duration of the experiment and are shown in {\bf c} and the inset of {\bf d}. {\bf c} One-half of the particles' (blue and green) blinking cycle, demonstrating that their optical axes are asynchronous. The incident electric field -- whose direction is set by the orientation of the polarizer (P) -- is de-polarized whenever the optical axis of the rotating particles is aligned with neither the polarizer nor analyzer (A). {\bf d} Computing the magnitude of the Fourier transform ($\sqrt{\mathcal{F}\mathcal{F}^*}$) of the blinking patterns (inset) of the two particles in {\bf b} shows that the frequencies at which the particles de-polarize the incident L.E.D. light are centered around $0.5\;\rm{Hz}$, corresponding to a rotation frequency of 0.125\;\rm{Hz}. The magnitude of the sum of transforms, $\left|\sum_i \mathcal{F}_i \right|^2$ (solid line), decays, confirming that the particles' orientations are out of phase.}  Scale bar: $5\;\mu\rm{m}$;}
\vspace{-0.75cm}
\label{fig:Async}
\end{figure}

%########### FIGURE2 ############
\begin{figure}[h!t]
    \centering
    \includegraphics[width=0.75\textwidth]{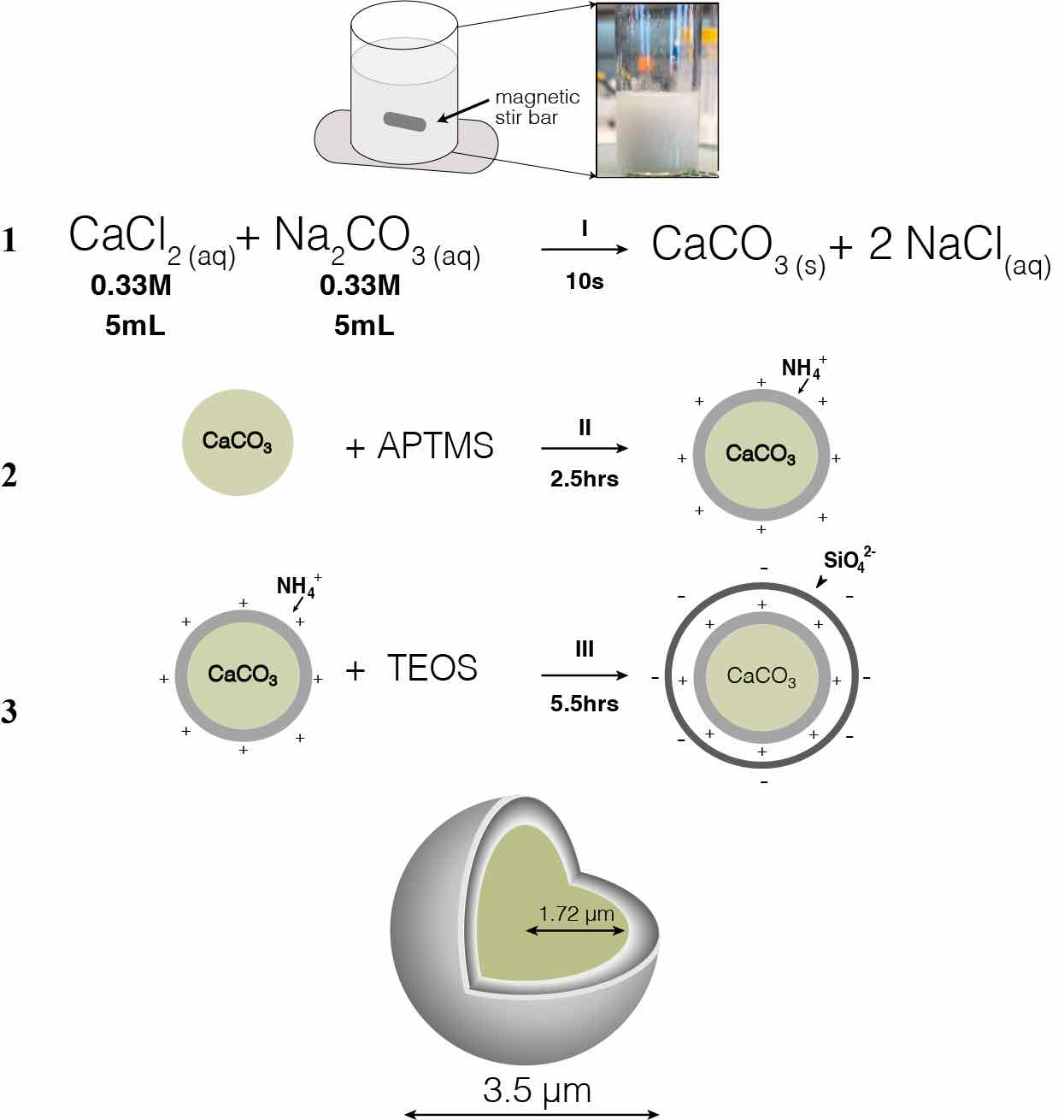}
    \caption{\textbf{Synthesis of silica-coated vaterite particles.} {\bf a} Synthesis procedure starting with \textit{(1)} mixing of equal volumes of buffered $\rm{CaCl_{2}}$ and $\rm{Na_{2} C.O._{3}}$ for 10s with a magnetic stir bar, followed by a two-step coating procedure starting with \textit{(2)} APTMS The $\rm{CaCl_{2}}$-A.T.P.M.S. solution was placed in a shaker for approximately 2.5 hours. Particles were then coated with \textit{(3)} TEOS and placed in a shaker for 5.5 hours. The two-step coating procedure was then repeated.}
    \label{fig:Synthesis}
\end{figure}

%########### FIGURE3 ############

\begin{figure}[!hb]
\centering
%\vspace{0.5cm} 
\includegraphics[width=0.65\textwidth]{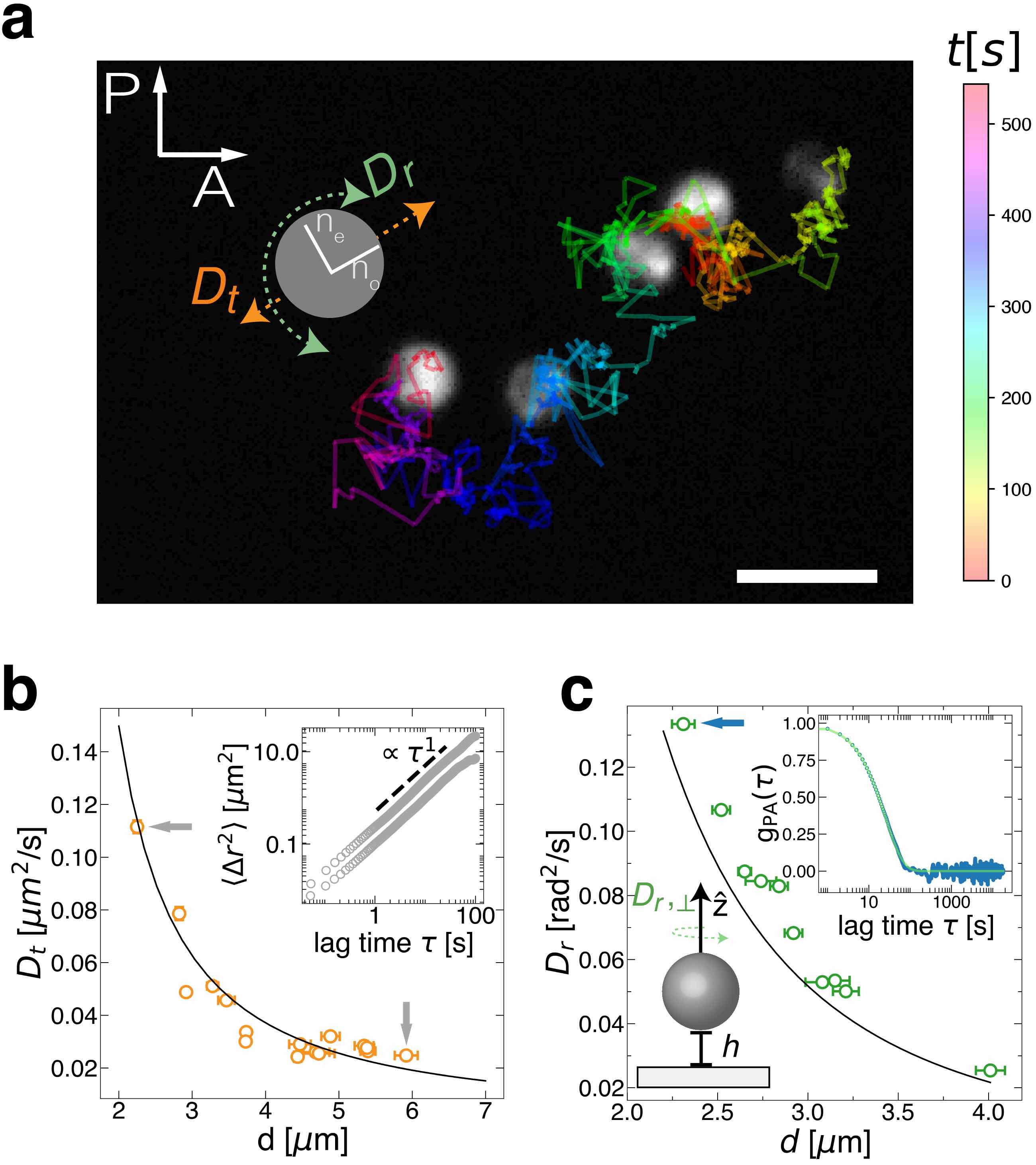} 
\caption{ \textbf{The translational and rotational diffusion of an individual particle is described by a sphere near a non-slip wall in Stokes flow}. {\bf a}   Snapshots (and time-colored trajectory) of a birefringent vaterite particle viewed through crossed-polarizers, showing Brownian translation and rotation. {\bf b} Near a no-slip wall (solid line), particles have lower translational diffusion $D_{t}$ relative to particles in bulk, consistent across a measured size range of $d=2-6\;\mu\rm{m}$.  \green{Inset shows the translational mean-squared displacement (MSDs) for particles with sizes $d = 1.92 \pm 0.11 \mu$m  and $ 5.92 \pm 0.35 \mu$m. The translational diffusion constants $D_{\rm{t}}$ obtained from these two MSDs are indicated by grey arrows in the main panel. {\bf c}  Rotational diffusion perpendicular to the wall, (spinning) $D_{\rm{r},\perp}$, measured using depolarization intensity decorrelation, $g_{\rm{PA}}$. $D_{\rm{r},\perp}$ is largely unchanged by the presence of a no-slip wall (solid line). The blue arrow indicates $D_{\rm{r},\perp}$ obtained from fitting $g_{\rm{PA}}$ for a $d =2.31 \pm 0.14 \mu$m particle (inset}). Scale bar: $5\;\mu\rm{m}$.}
\label{fig:TransDyn}
\end{figure}

%########### FIGURE4 ############

\begin{figure*}[h!]
\centering
%\vspace{0.5cm} 
\includegraphics[height=0.7\textwidth]{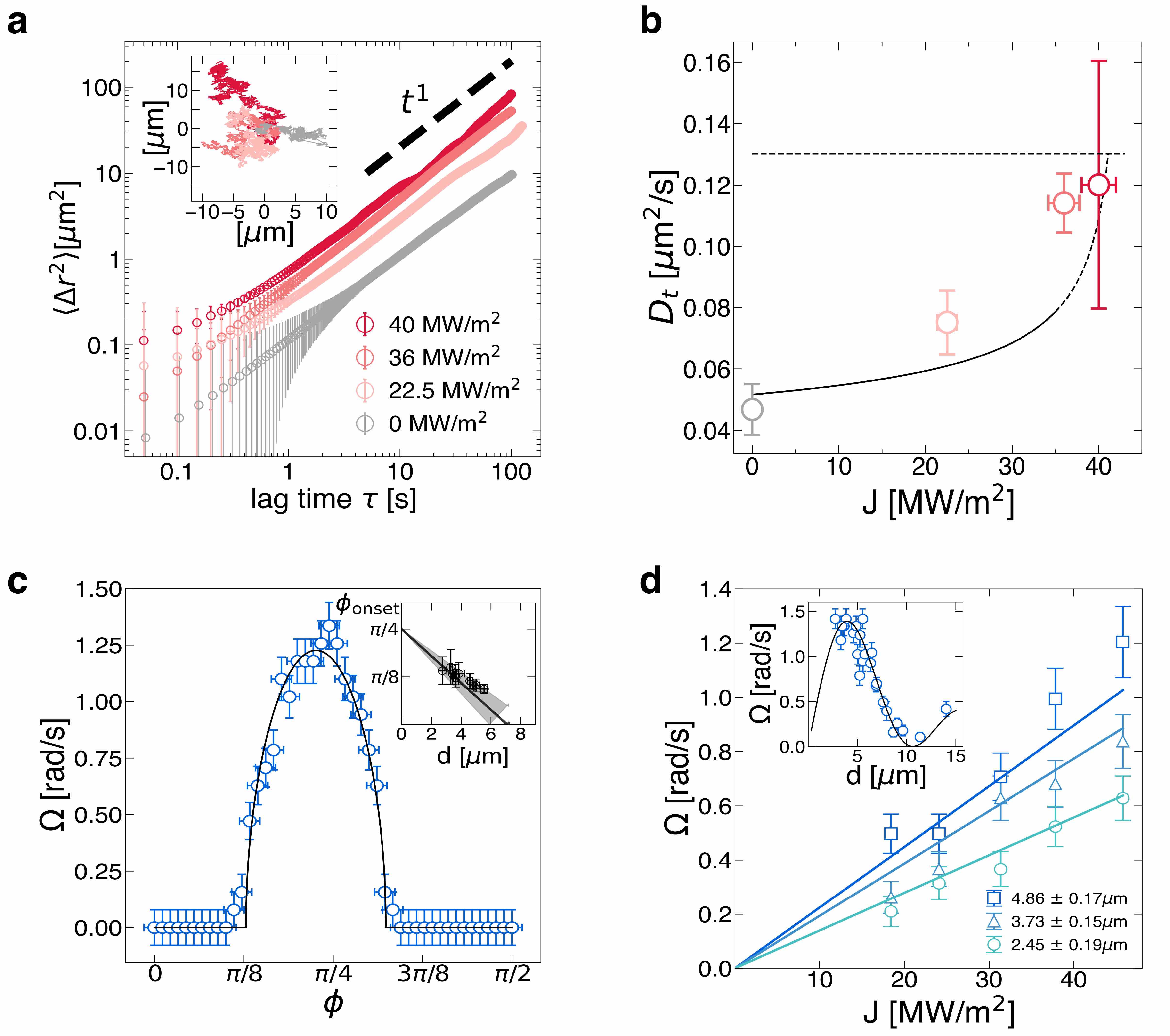} 
\caption{\textbf{Measurement of rotational and translational dynamics in a uniform optical torque field.} Individual rotors freely diffuse and spin in a force-free optical torque field. {\bf a} The ensemble-averaged mean square displacement $\langle \Delta r^2\rangle$ is linear (diffusive) for different photonic fluxes $J$. A representative trajectory of a single particle for different photonic fluxes is shown in the inset. {\bf b} Translational diffusion increases with $J$ as the gravitational height $h_{g}$ increases (solid line) until it approaches the bulk value (dashed line) {\bf c} The effective birefringence of a particle, $\Delta n$, can be measured by monitoring the minimal ellipticity, $\phi$, where rotation begins for different particle sizes (inset). {\bf d} Measured spin rates for different fluxes and different particle sizes (inset) as predicted by Eq. \ref{eqRotFreq} (solid lines), are consistent with measured birefringence.}
    \label{fig:RotDyn}
\end{figure*}

%########### FIGURE 5 ############
\begin{figure}[h!t]
    \centering
    \includegraphics[width=0.75\textwidth]{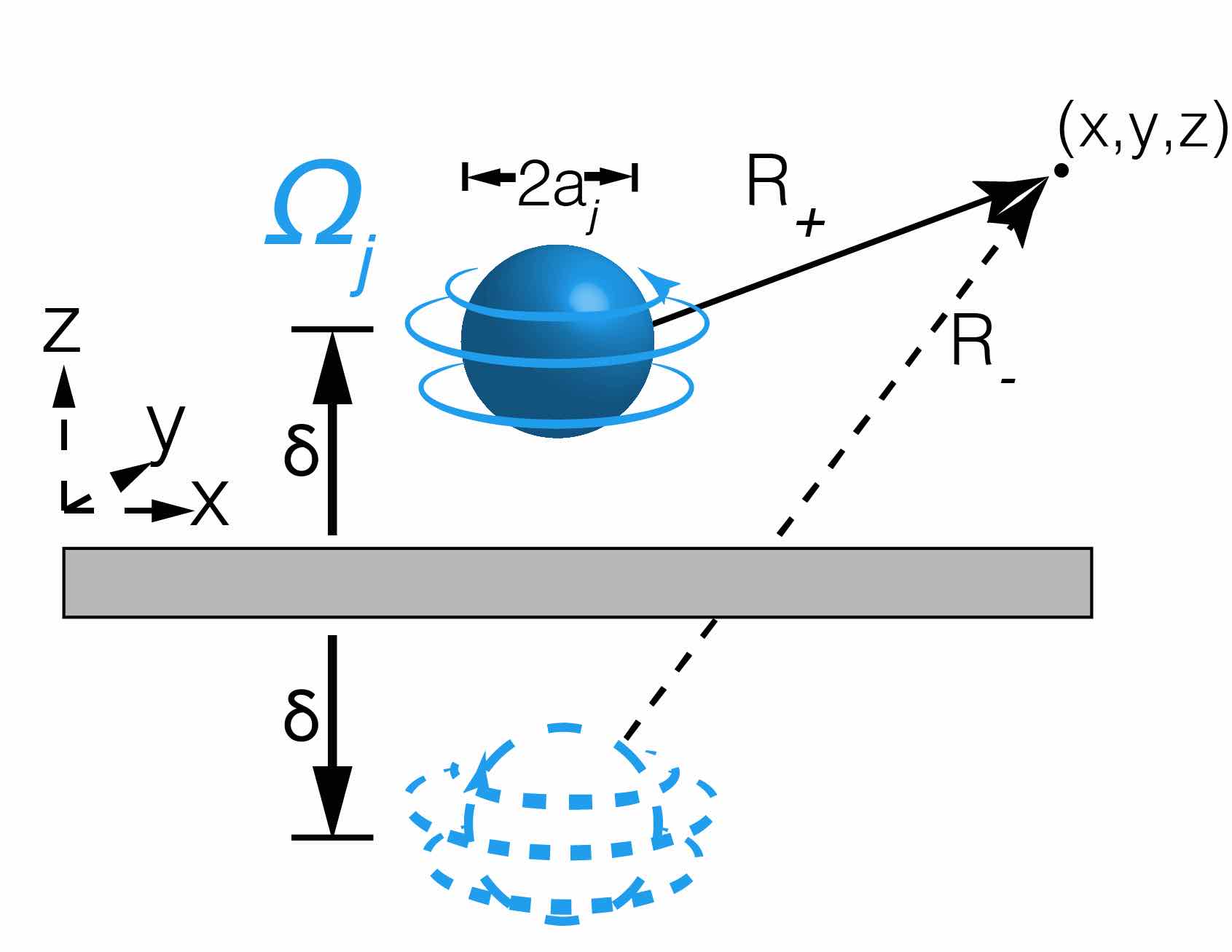}
    \caption{\textbf{Flow field of a single rotor spinning near a wall.} Geometry of a rotating sphere and its corresponding image charge generated near a solid no-slip wall while rotating about an axis perpendicular to the wall.}
    \label{fig:SingleRotor}
\end{figure}
%\end{singlespace}

%########### FIGURE6 ############
%\begin{singlespace}

\begin{figure*}[h!t]
\centering
\includegraphics[width=.75\textwidth]{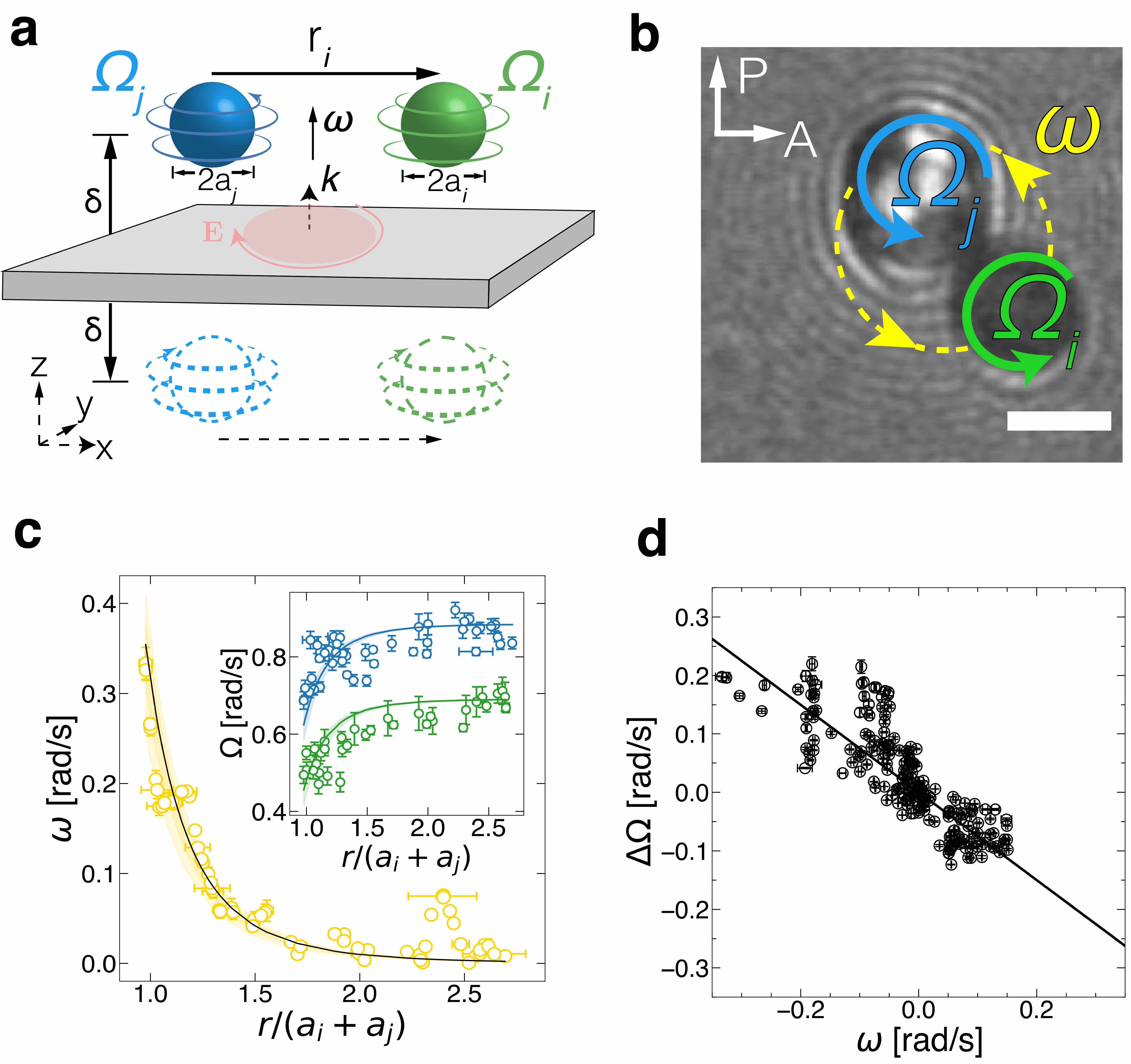} 
\caption{\textbf{Pairs of free rotors couple hydrodynamically through rotation and translation.} {\bf a} Diagram of an orbiting pair near a wall with corresponding image charges. {\bf b} Snapshot of two mutually advecting particles.  {\bf c}
\green {Angular speed $\omega$ of two orbiting spheres (diameters $6.7\;\mu$m and $5.1\;\mu$m) at different separations, along with the dependence of the spinning rate $\Omega$ on the normalized separation (inset)}. \green{Curves show the predicted spinning rates for each particle at different separations -- derived from Faxen's laws (Eqs. 6 and 7 in the main text) -- given the particles' asymptotic spinning rate, $\Omega_{i,j}^0$, measured at large separations} (colors correspond to panel {\bf b}). {\bf d} The change in spin $\Delta \Omega$, and the orbital frequency $\omega$ for rotors of varying size, separation, and optical flux follows a geometric relation given by Eq.~\ref{eqSpinOrbit}, where $\alpha = 4$ (solid line). \green{The scatter relative to the trend line originates from thermal fluctuations in transient orbits of freely diffusing Brownian rotors (see Supplementary
Information)}. Scale bar:  $5\;\mu\rm{m}$.}
\label{fig:SpinOrbit}
\end{figure*}

%%%%%%%%%%%%%%%%%%%%%%%%%%%%%%%%%%%%%%%%%%%%%%%
%%%%%%%%%%%%%%%%%%%%%%%%%%%%%%%%%%%%%%%%%%%%%%%
%%%%%%%%%%%%%%%%%%%%%%%%%%%%%%%%%%%%%%%%%%%%%%%

\clearpage

\clearpage

\renewcommand{\theequation}{S\arabic{equation}}
\setcounter{equation}{0}

\renewcommand{\thefigure}{S\arabic{figure}}
\setcounter{figure}{0}
\begin{mytitlepage}
    
\title{Hydrodynamic spin-orbit coupling in asynchronous optically driven micro-rotors\\ Supplementary Information}
\maketitle
\vspace{-15mm}

\section*{Optical setup}
Our instrument is designed to apply a uniform optical torque field by delivering a collimated beam to the sample plane. A schematic diagram of the setup can be seen in Supplementary Fig.\hspace{-1mm}~\ref{fig:optics}a. The experimental system consists of a 3mm infrared (IR) laser, $\lambda=1064$nm (IPG Photonics, $M^{2}=1$), propagating first through a half-wave plate ($\lambda/2$) followed by a Galilean beam contractor. The telescope can shrink the beam by a factor of 30, while preserving the initial collimation of the laser source. Once the beam diameter has been reduced, the laser propagates through a polarizing beam-splitter (PBS) cube where the S-component of light is reflected upwards. The handedness of the linearly polarized light can be altered via a quarter wave plate ($\lambda/4$) fixed immediately after the PBS atop a precision manual rotation mount. 

We image the distribution of the incident near-infrared beam using an infrared viewing card. The pixel intensities of the image are then fit to a Gaussian surface of revolution to obtain an intensity distribution of the beam flux at the sample plane. The flux distribution $J\left(\bm{r}\right)$ can be written as 
\begin{equation}
J\left(\bm{r}\right)=J_0 \exp\left[-2\frac{\textbf{r}^{2}}{\sigma^2}\right]
\end{equation}
The prefactor corresponds to the flux at the center of the beam and is obtained by integrating the 2-dimensional Gaussian over the area of the beam.
Supplementary Fig.\hspace{-1mm}~\ref{fig:optics}b shows a typical $J\left(\bm{r}\right)$ for $\sigma=372.8 \pm 5.6 \mu$m (FWHM = $438.9 \pm 6.6\mu$m). We can adjust the flux at the sample by changing the size of the incoming beam, modulating the power at the laser head, or tuning the orientation of the $\lambda/2$- plate.

We illuminate the sample using a 505 nm LED (Thorlabs) to image our photonic rotors. The light emitted by the LED passes through a N-BK7 ground glass diffuser (600 grit) and is then pseudo-collimated with an N-BK7 plano-convex lens (f=25.4mm) that acts as a primary collector lens. After the plano-convex lens, a secondary achromatic doublet lens (f=76.2mm) forms an image of the filament at the position of a field diaphragm. A condenser lens focuses the light at the objective's back focal plane. Particles are imaged using an infinity-corrected Leica HCX PL APO 40x objective (NA=0.85). High-pass index matched IR filters mounted before the camera are used to prevent interference of IR light with the 3.1MP monochrome camera (Imaging Source). 

\begin{figure}[h!]
    \centering
    \includegraphics[width=10cm]{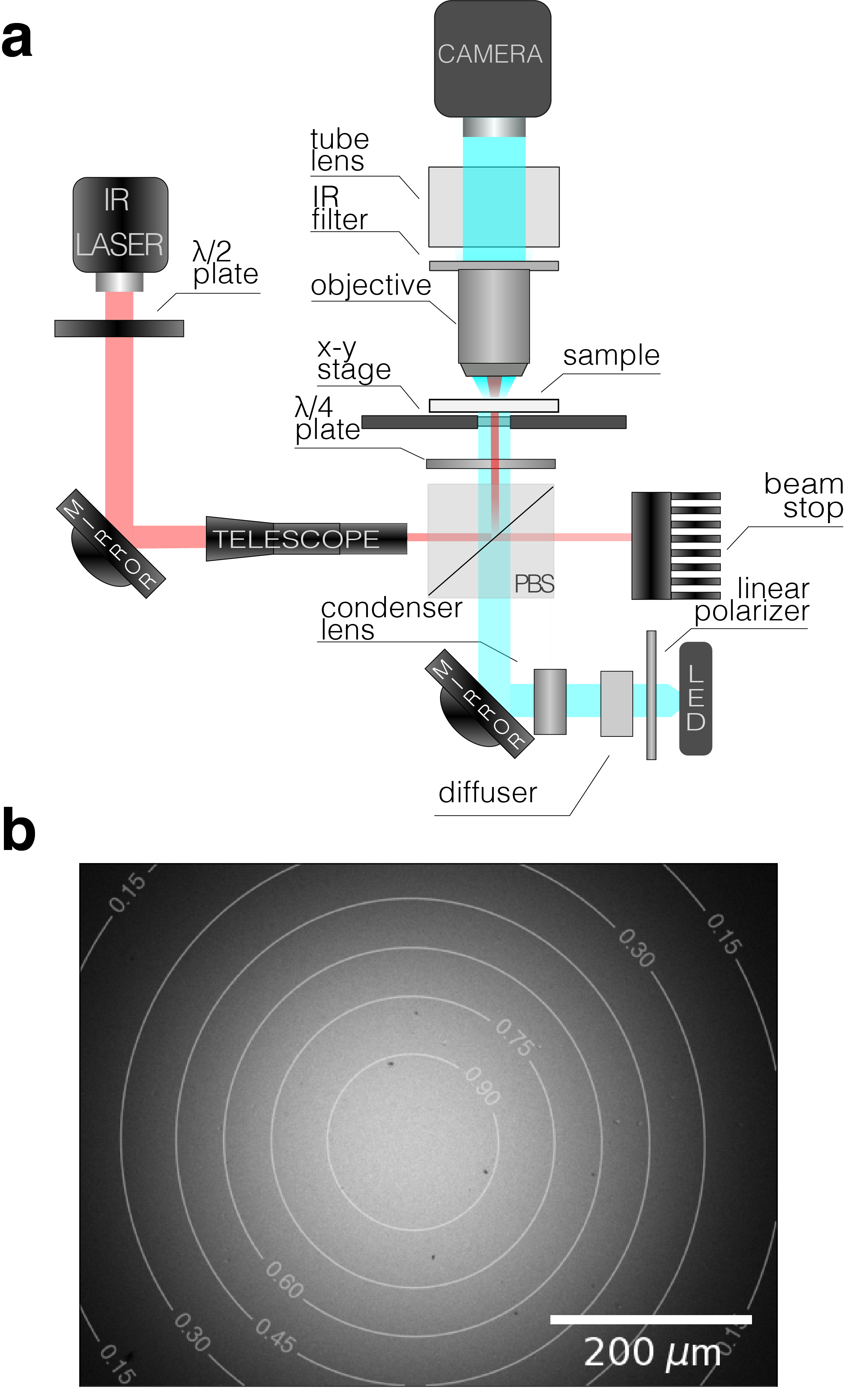}
    \caption{\textbf{Optical setup used to create a torque field to drive birefringent micro-rotors.} {\bf a} Detailed schematic of the optical elements used to couple the imaging and laser systems to deliver a defocused beam to the sample plane. {\bf b} Broad field image of the laser beam (contours show relative power given by Gaussian fit). Experiments were performed in a region $\sim 165\mu$m x $125\mu$m, where gradients in the flux are negligible.}
    \label{fig:optics}
\end{figure}

\section*{Measuring rotational diffusion using transmission decorrelation through crossed-polarizers }
Rotational diffusion is measured by calculating the auto-correlation of the depolarization of the light intensity transmitted through the particles, as measured using cross-polarizers on our custom-built microscope (see Supplementary Fig. 1(a)). Following previous work~\cite{Berne2000, Lisicki2014}, we define the intensity auto-correlation function as $g_{PA}(\tau)=\langle I_{PA}^{*}(0)I_{PA}(\tau)\rangle/\langle|I_{PA}|^{2}\rangle$ where $I_{PA}$ is the time-averaged scattered intensity field detected by the camera, and $\tau$ is the lag time. To leading order, the translational and rotational dynamics are uncoupled, thus enabling direct measurement of the diffusion matrix $\mathbf{D}_{r}$ by fitting $g_{PA}=\exp(-6\mathbf{D}_{r}\tau)$  (Fig. 3 in the main text). In the lubrication limit, diffusive rolling, $D_{r,||}$ (rotation on an axis parallel to the wall) is much slower than diffusive spinning (rotation perpendicular to the wall $D_{r,\perp}$) which dominates the diffusion matrix,  $\mathbf{D}_{r} \approx D_{r,\perp}$~\cite{Lisicki2014}.
\section*{Measuring the translational diffusion constant of rotors}

\green{We record videos of freely-diffusing vaterite micro-spheres and track their instantaneous positions of particles using the Python package TrackPy. TrackPy was downloaded and used without further modification \cite{allan_daniel_2016_60550}. The package implements the widely used Crocker-Grier particle tracking algorithm \cite{crocker_methods_1996}, which identifies local brightness maxima within an image as candidate particle locations. We track particles diffusing throughout the movie and extract their diffusion constants by computing their mean squared displacements, $\langle \Delta r ^2\rangle = 4D_{t}\tau$ for different lag-times $\tau$. The y-intercept of a linear fit in log directly measures $D_{t}$.}
\section*{Measuring the reflection coefficient of a vaterite rotor}
\subsection*{Quantifying the back-scattered radiation by a photonic rotor}
Back-scattered light from a photonic rotor generates a radiation pressure that counteracts the gravitational force $F_{g}=\pi \Delta \rho g d{^3}/6$ (Supplementary Fig.\hspace{-1mm}~\ref{fig:radiationpressure}). For a given flux $J$, the average force on a particle with diameter $d$ is given by $\langle F_{rad}\rangle=\frac{R J\pi d^2}{4 c}.$
Here $c$ is the speed of light, and $R$ is the reflection coefficient. For vaterite micro-spheres, absorption of IR light is negligible \cite{Arita2016}. Micro-spheres travel at a constant velocity through the $100 \mu$m capillary when $\langle F_{rad}\rangle>F_{g}$. This upward motion is balanced by the viscous drag force $F_D = 3\pi\eta d v$. Here $v$ is the average rise velocity of the particle as it travels through a $100\mu$m tall capillary. \green{The procedure for measuring $v$ can be found in the subsection below.}

Balancing the forces acting on the particle gives,  
\begin{equation}
\frac {R J_{0} }{12c \eta}- \frac{1}{18}\frac{\Delta \rho g}{\eta}d = \frac{v}{d},
\label{radiationPressure}
\end{equation} 
where the buoyant density $\Delta \rho$ = 1.43 g/cc for vaterite particles suspended in $\rm{D_{2}O}$.
The velocities for different particle sizes and fluxes are shown in Supplementary Fig.~\ref{fig:radiationpressure}. $R$ is extracted by linear fit of  Eq.~\ref{radiationPressure} to the data. We find that $R=0.22\pm 0.01$ -- to our knowledge, this is the first direct measurement of the reflection coefficient of vaterite micro-spheres and is in good agreement with spherical particles of similar indices of refraction \cite{Ashkin1970}. 
\green{\subsection*{Measuring the average rise velocity $v$ of a particle in the presence of an optical flux}
When $\langle F_{rad}\rangle > F_{g}$, particles steadily rise. To monitor the out-of-plane motion of the particles, we first focus on the particles as they freely diffuse in the plane in the presence of no optical flux. Using a translation stage, we vertically shift the imaging focal plane by $100 \mu$m, corresponding to the capillary's height where the particles are suspended. We time how long it takes for the particles to re-appear in focus at the top of the capillary. The depth of field of our 40x objective is $\approx 1 \mu$m -- less than the diameter of the particles -- ensuring good accuracy of our measurements. 
}

\begin{figure}[h!t]
    \centering
    \includegraphics[width=10cm]{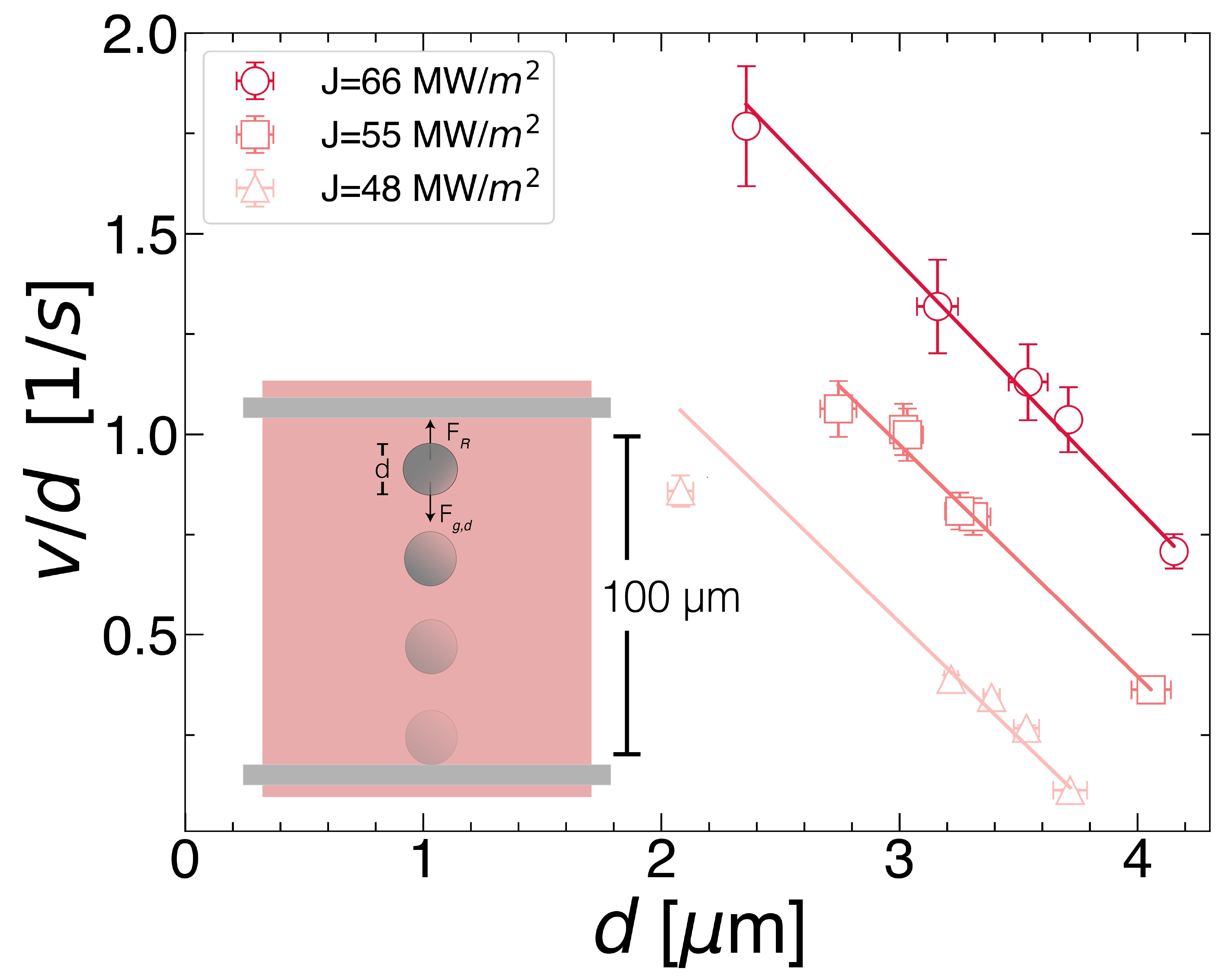}
    \caption{\textbf{Linear fits to the ratio between the average rise velocity and particle size $v/d$ allows for direct measurement of the reflection coefficient R for vaterite microspheres.} Across three different fluxes, we compute an average $R=0.22 \pm0.01$. Buoyant densities obtained from the slope of the lines vary by $<10\%$ from the expected value.}
    \label{fig:radiationpressure}
\end{figure}

\green{\section*{Spinning angular frequency $\Omega$ of a birefringent particle}
In the low-Reynolds number limit, a sphere spinning in a fluid at angular frequency $\Omega$ experienced a viscous torque $\tau_{\rm{v}} = \pi\eta d^3\Omega$, dependant on the viscosity of the fluid $\eta$ and the particle's diameter, $d$. $\tau_{\rm{v}}$ is balanced by the optical torque, $\tau_{\rm{p}}$, that is responsible for generating the spinning motion. The magnitude of $\tau_{\rm{p}}$ depends on the incident light's polarization ellipticity $\phi$ and wavelength $\lambda$ according to \cite{Friese1998},
\begin{equation}
    \tau_{\rm{p}} = \frac{\lambda P T}{2\pi c}\left[\left(1-\cos\frac{2\pi\Delta nt}{\lambda}\right)\right]\sin2\phi - \sin\left(\frac{2\pi\Delta nt}{\lambda}\right)\cos2\phi\sin2\theta.
    \label{spinAlign} 
\end{equation} 

Here, $P$ represents the power incident on the particle, and $c$ is the speed of light. The first term of Eq.~\ref{spinAlign} corresponds to the "spinning torque"; the argument of $\cos$ can be thought of as the degree to which the particle acts as a phase-retarder, providing the maximum change in the transmitted light's spin-angular momentum when the particle is exactly the thickness of a half-wave plate. Therefore, it must depend on the physical properties of the particle, such as its birefringence $\Delta n$, transmissivity $T$, and thickness $t$. The second term in Eq.~\ref{spinAlign} is a torque that seeks to align the fast axis of the particle with the plane of polarization.
To derive an equation for the spinning rate in the presence of a circularly polarized incident beam $(\phi = \pi/4)$, we approximate the spherical particles as cylinders so that their diameter $d\approx t$ \cite{Vaippully2020}. In this limit, the alignment torque vanishes, and the optical torque is proportional to $\Delta \sigma \equiv 1-\cos\left(\frac{2\pi\Delta n d}{\lambda}\right)$. Rewriting $P$ in terms of the power flux $J$, $P = J\pi d^2/4$, the optical torque becomes $\tau_{\rm{p}} = \Delta \sigma T J\lambda d^2/8c$. Equating $\tau_{\rm{p}}$ and $\tau_{\rm{v}}$ and solving for $\Omega$ gives,
 \begin{equation}
        \Omega=\frac{\rm{T} J \lambda}{8\pi c\eta d}\left[1-\cos\left(\frac{2\pi\Delta n d}{\lambda}\right)\right]. 
        \label{eqRotFreq} 
    \end{equation}  
}

\begin{figure}[h!]
    \centering
    \includegraphics[width=15cm]{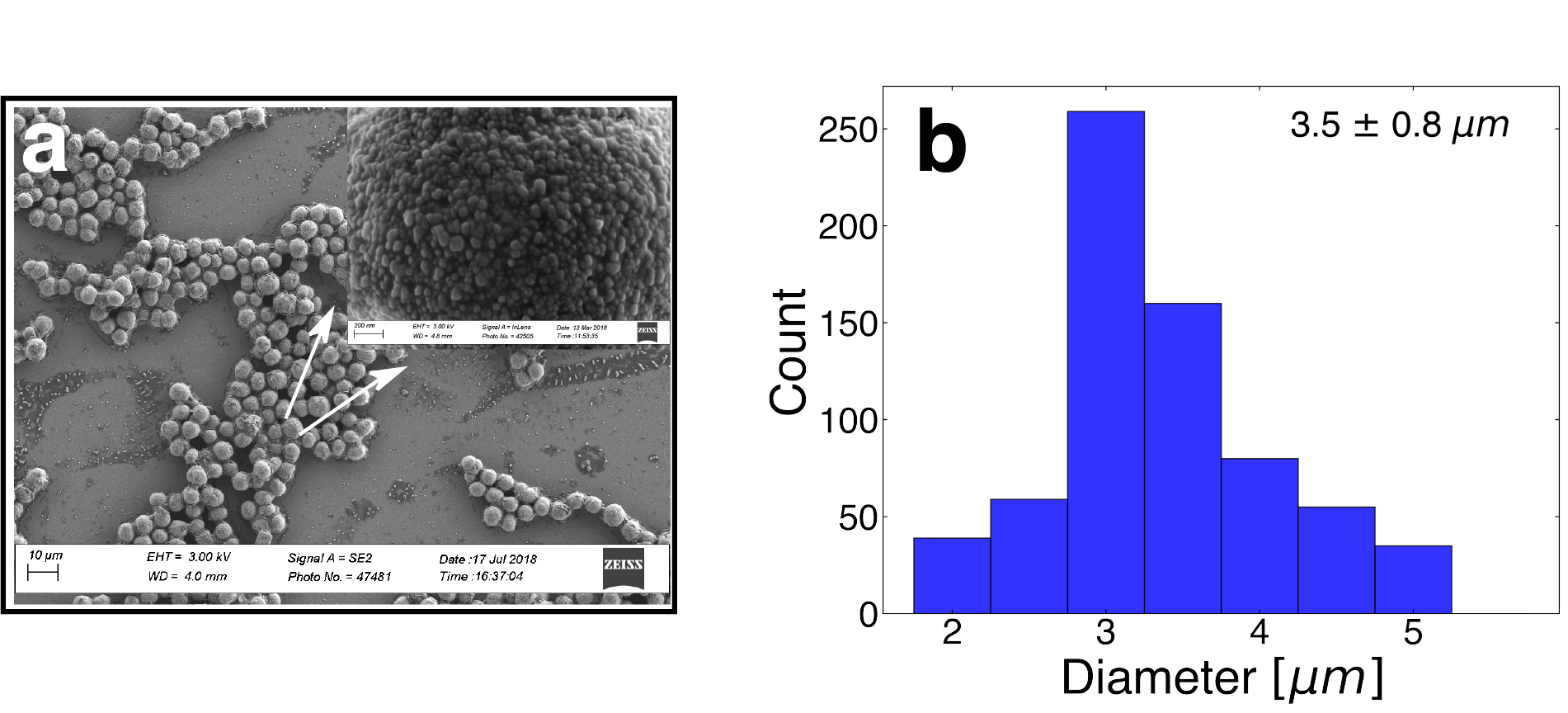}
    \caption{\textbf{Characterization of synthesized vaterite micro-spheres.} {\bf a} SEM micrograph of vaterite microspheres. Inset shows the poly-crystalline structure of the particle's surface. {\bf b} Size distribution of vaterite microspheres.}
    \label{fig:synthesis_2}
\end{figure}
\section*{Measuring the frequency and phase of a rotating particle}
When spinning, the angle $\theta$ a vaterite particle's optical axis makes with incident linearly polarized light varies periodically so that a particle's rotation frequency $f\propto \dot{\theta}$. Throughout one period, vaterite particles depolarize the transmitted light from the polarized LED source four times. The observed depolarizations, or "blinks", occur whenever the optical axis of the particle coincides with neither the axis of the polarizer nor the analyzer (Fig.~1c). Tracking the transmitted light intensities of individual particles as a function of time, $I_{i}\left(t\right)=I_{0}\sin^2\left(2f_{i}t+\Phi_{i}\right)$, allows for a direct measure of each particles' rotation frequency and phase, $\Phi_{i}$. We normalized all signals so that $I_0 =1$.

To measure $\Phi_{i}$ and $f_{i}$ of rotating particles, we compute the Fourier transform of their transmitted light intensities
\begin{equation}
\begin{aligned}
\mathcal{F}_i=\mathcal{F}_{i}\left[I_{i}\left(t\right)\right]\left(\omega\right)&\equiv \int dt e^{-\mathrm{i} \omega t} I_{i}\left(t\right)\\
& \propto -I_{0}e^{2\mathrm{i}\Phi_{i}}\delta\left(\omega - 4f\right),
\end{aligned}
\label{fourier}
\end{equation} where we have assumed that $\omega,\Phi_i, f$ are positive quantities. Eq. \ref{fourier} allows us to compare the phases of multiple particles globally. For individual particles, the magnitude of the Fourier transform $\sqrt{\mathcal{F}_{i}\mathcal{F}_{i}^{*}} \propto \delta\left(\omega-4f\right)$. However, the sum of N-individual Fourier transforms, $F_{N} = \left|\frac{1}{N}\sum_i^{N} \mathcal{F}_i \right|^2$, decays with the number of particles. Explicitly, 
\begin{equation}
\begin{aligned}
F_{N} &=  \left|\frac{1}{N}\sum_i^{N} \mathcal{F}_{i}\left[I_{i}\left(t\right)\right]\left(\omega\right) \right|^2\\
&=\left|\frac{1}{N}\sum_i^N \int dt e^{-\mathrm{i} \omega t} I_{i}\left(t\right) 
\right|^2\\
&\propto \frac{I_{0}^{2}}{N^2} \delta^2\left(\omega-4f\right)\left[N+\sum_i^N \sum_k^{N} e^{2\mathrm{i}\left(\Phi_i-\Phi_k\right)}\right].
\end{aligned}
\label{fourierExplicit}
\end{equation}
As an example, let us consider the case of $N=2$ particles, Eq. \ref{fourierExplicit} becomes, 
\begin{equation}
   F_{N} = \frac{I_0^{2}}{4}\delta^2\left(\omega-4f\right)\left(2+e^{2i\left(\Phi_{1}-\Phi_{2}\right)}+ e^{2i\left(\Phi_{2}-\Phi_{1}\right)} \right)
\end{equation}
When particles have the same global phases, $\Phi_{1} = \Phi_{2}$, $\sqrt{F_{2}F_{2}{^\star}} = I_{0} = 1$. When the optical axes of the particles are perfectly out of phase $\Phi_{1}-\Phi_{2} = \pi/2$, $\sqrt{F_{2}F_{2}{^\star}} =0$. For all other intermediate relative phases, as -- is the case in Fig. 1(c,d) -- the asynchronous phases of the light intensities do not add up constructively, and the amplitude of the sum of the individual Fourier transforms is reduced.

\section*{Flow generated by a rotating sphere near a wall in the Stokes-flow regime}

% \begin{figure}[!h]
%     \centering
%     \includegraphics[width=10cm]{SI_fig_imageCharge.png}
%     \caption{Geometry of a rotating sphere and its corresponding image charge generated near a solid no-slip wall.}
%     \label{fig:geometry}
% \end{figure}

Consider an isolated sphere with radius $a$
subjected to a constant torque $\bm{\tau^0}=8\pi\eta a^{3}\mathbf{\Omega}_{j}^{0}$, where $\eta$ is the fluid viscosity, and $\Omega^{0}_{j}$ is the initial angular speed of the rotor in the absence of any neighboring rotating particles. The Green's function for this
rotlet is \cite{Blake1974}
\[
\mathbf{G_{ij}}=\frac{\epsilon_{ijk}r_{k}}{r^{3}}
\]
and the corresponding free space flow is 
\[
\mathbf{u_{j}\left(r\right)}=\epsilon_{ijk}\Omega_{j}^{0}a^{3}\frac{r_{k}}{r^{3}}.
\]
Here $\epsilon_{ijk}$ is the Levi-Cevita symbol with indices representing the $x,y,z$ directions in the Cartesian coordinate system, and $r$ is a three-dimensional (3D) vector, $|r|\equiv\sqrt{x^{2}+y^{2}+z^{2}}$, from the source of the disturbance to a point $\left(x,y,z\right)$ in space. 

In our experiments, rotors spin near the wall of a capillary. We consider a  rotating sphere spinning about an axis perpendicular to a nearby plane located at $z=0$. The sphere's center is at a height $\delta = h + a$ (see Fig.~5 in the main text). We assume that the plane has a no-slip boundary condition, $\left(\bm{u_j}(z=0)=0\right)$, so that we may approximate the far-field fluid flow
as the superposition of two rotlets with equal and opposite torques,$\bm{\tau}=\pm\bm{\tau^0},$ located at $z=\pm\delta$, respectively (Fig.5).
The resulting fluid flow is,  
\begin{equation}
    \bm{u}_{j}\left(r\right) =\Omega_{j}^{0}a^{3}\left[\left(\frac{1}{|R_{+}|^{3}}-\frac{1}{|R_{-}|^{3}}\right)\left(-y\mathbf{\hat{x}}+x\mathbf{\hat{y}}\right)\right]
    \label{flowFieldFull}
\end{equation}

Here, $|R_{\pm}|\equiv \left(x^2+y^2+\left(z\mp\delta\right)^{2}\right)^\frac{1}{2}$ representing the distance to a point $\left(x,y,z\right)$ in space from the source and image charges, respectively. Next, we consider specifically the far-field limit $\left(\delta\ll r\right)$ where $|R_{\pm}|=\left[( r^{2}\left(1\mp\frac{2\delta z}{r^{2}}\right)\right]^{\frac{1}{2}}$ and $|R_{\pm}|^{-3}\approx\frac{1}{r^{3}}\left(1\pm\frac{3\delta^2}{r^{2}}\right)$. In the final equality, we make the approximation that the flow is strictly two-dimensional (2D) and confined to the special symmetry plane located approximately one particle radius from the wall at $z=\delta$. Eq.\ref{flowFieldFull} reduces to:

\begin{equation}
    \bm {u}_{j}\left(r\right) = \frac{6\Omega_{j}^{0}a^{3} \delta^2}{r^4}\bm{\hat{\theta}}.
    \label{approxFlow}
\end{equation}
Note that the $r^{-4}$ scaling is a result of the conversion from Cartesian (with basis vector $\bm{\hat{x}},\bm{\hat{y}}$) to polar coordinates (with basis vectors $\bm{\hat{r}},\bm{\hat{\theta}}$). The polar component of the flow, $\bm{\hat{\theta}}$, has been rewritten using $-r\sin\theta\bm{\hat{x}}+r\cos\theta\bm{\hat{y}}=-y\bm{\hat{x}}+x\bm{\hat{y}} = r\bm{\hat{\theta}}$. In Eq. \ref{approxFlow}, $r$ is now the distance in the 2D plane.

Faxen's first law \cite{Happel1983},  $\bm{F}_i = 6\pi \eta a\left\{\left[\bm{u}_{j}\left(\bm{r}\right)+\frac{1}{6}a^2\nabla ^2 \bm{u}_j\left(\bm{r}\right)\right]_{\bm{r}=\bm{r}_i}-\bm{v}_i\right\}$, in the presence of no external forces ($\bm{F}_i=0$) reduces to
\begin{equation}
\mathbf{v}_{i}=\frac{d\mathbf{r}_{i}}{dt}=\bm{u}_{j}\left(\bm{r}\right)+\mathcal{O}\left(\frac{1}{r^{6}}\right)
\label{advection}
\end{equation}

Eq. \ref{advection} describes the tangential velocity of particle $i$ resulting from the advective flow field generated by particle $j$. The modification of particle $i\rm{'s}$ angular velocity (spinning rate) due to the vorticity of $j\rm{'s}$ flow is found by application of Faxen's second law\cite{Happel1983}, 

\begin{equation}
\bm{\Delta \Omega}_i \equiv \bm{\Omega}_i - \bm{\Omega}^0_i = \frac{1}{2}\nabla \times \bm{u}_j\left(\bm{\bm{r}=\bm{r}_i}\right). 
\end{equation} 
Rewriting Eq.\ref{approxFlow} in terms of a general flow power law, $\alpha$ and prefactor, $A$, $\mathbf{u}_{j}\left(r\right)=Ar^{-\alpha}{\Omega}^{0}_{j}\bm{\hat{\theta}}$. We compute the curl as $\left(\nabla\times \mathbf{u}_{j}\right) =  \frac{1}{r}\left(\frac{\partial}{\partial r}\left[Ar\cdot r^{\alpha}\Omega^{0}_{j}\right]\right)$ 
$=\mathbf{u}_{j}\left(r\right)\left(1-\alpha\right)/r$. Considering the case where rotors are of equal size, we arrive at 
\begin{equation}
    \Delta\mathbf{\Omega}=\frac{1-\alpha}{4}\bm{\omega},
    \label{spinOrbit}
\end{equation}
where we have rewritten the expression in terms of the orbital frequency of the rotating pair about their common center by using 
$\bm{\omega}=2\mathbf{\omega}_{i}=2\mathbf{u}_{j}\left(r\right)/r$. 
For the case of a particle near a wall, $\alpha=4$ and Eq.\ref{spinOrbit} 
reduces to $\Delta\bm{\Omega}=-3\bm{\omega}/4$.

\subsection*{Measuring the change in the instantaneous spinning rate $\Delta\Omega$ of a rotor}

We record videos of hydrodynamically-coupled rotors and track their trajectories and transmitted light intensities $I_{i}\left(t\right)$ using TrackPy \cite{allan_daniel_2016_60550}. For each rotor pair, local maxima and minima in $I_{i}\left(t\right)$ are found based on a topographic prominence algorithm distributed through the \textit{SciPy} library. We compute the difference between neighboring minima and maxima in $I_{i}\left(t\right)$, proportional to $1/8$ of the particle rotation frequency. For a given window, the instantaneous $\Omega$ is found from the mean and standard error of the minima-maxima difference. The window is then advanced by a $1/8$ of a rotation period, and this procedure is repeated until the entire signal is evaluated.

\subsection*{Expected fluctuations in the extracted spin rate for measuring the spin-orbit coupling}

The accuracy of measuring the instantaneous rotation rate of a Brownian rotor is limited by the time over which the orientation rate is tracked. An active rotor that follows the overdamped Langevin equation $\dot{\theta} = \Omega_0+\eta\left(t\right)$ (where $\Omega_{0}$ is the nominal spinning rate, and $\eta\left(t\right)$ is the thermal noise term), follows the average dynamic analogs to a Brownian particle undergoing thermal diffusion in 1D:

\begin{equation}
\begin{split}
    &\langle \Delta \theta \rangle = \Omega_0 t\\
    &\langle \Delta \theta^2\rangle = 2D_r t.
\end{split}
\label{eqBrownianRotor}
\end{equation}

The mean evolution of the orientation ($\Delta \theta \equiv \theta \left(t\right) -\theta\left(0\right)$) has a drift-like term  ($\langle \Delta \theta \rangle ^2 \propto \left(\Omega_0 t\right)^2$). When tracking the change in orientation over some finite time, $\tau$, the instantaneous spinning rate can be approximated by the mean  $\langle\Omega_0\rangle \approx \langle \Delta \theta\rangle /\tau$. Measurement over a finite time will also be subjected to thermal fluctuations, and hence the measured instantaneous spinning rate will have a finite uncertainty: $\langle \Delta \Omega _0\rangle \approx \sqrt{\langle \Delta \theta ^2\rangle}/\tau \approx \sqrt{2D_r/ \tau}$.  Therefor there is an intrinsic uncertainty for the measured spinning rate when evaluated over a finite period of time as done in an experiment. The relative error decreases with time as: ${\bf \rm{RE}} = \frac{\langle \Delta \Omega _0\rangle}{\langle \Omega _0\rangle} = \frac{\sqrt{2D_r}}{{\Omega_0\sqrt{\tau}}}$.

When two optical rotors interact, they make a transient orbit, until they diffuse apart. We evaluated the instantaneous spinning rate over the duration $\tau_B$ --- the time between two blinking events (see Fig. 1c in the main text). We can now compute the uncertainty in the spinning rate for a typical $4\;\mu\rm{m}$ particle using the empirically found parameters: $D_r\approx 0.02\;\rm{rad}^2/\rm{s}$, $\Omega_0 \approx 1\;\rm{rad}/\rm{s}$, $\tau_B\approx 2\;\rm{s}$ (see Fig. 1c, Fig 3c, and Fig. 4d in the main text). This gives relative error of ${\bf \rm{RE}}\approx 0.14$. Compared with the nominal spinning rate, the relative error is not very large. But when compared with $\Delta \Omega$ -- that is the \textit{change }in the spinning rate (originating from the hydrodynamic spin-orbit coupling) -- this thermal fluctuation is more significant. We measure a relative change in spinning rate $\Delta \Omega /\Omega_0 \lessapprox 0.3$. This value is consistent with previously theoretically predicted value~\cite{Davis1969}. As can be seen in Fig. 6d, the thermal fluctuations are small (but compatible) to the measured spin-orbit coupling.

\end{mytitlepage}

% %%%%%%%%%%%%%%%% Videos %%%%%%%%%%%
% % 1. Best hundreds of rotors movie. Selected few (3-5) also show trajectory.
% % 2. Best spin orbit movie, showing trajectory and playing the intensity (sinus) plot.
% % 3. Best movie of a freely diffusing particle showing blinking.
% % 4. Best single rotor movie.

%################### Bibliography ####################
\bibliographystyle{naturemag}
\bibliography{./opticalRotors_nat}

\end{document}